\documentclass[pdflatex,sn-mathphys-num]{sn-jnl}% Math and Physical Sciences Numbered Reference Style
\usepackage{graphicx}%
\usepackage{multirow}%
\usepackage{amsmath,amssymb,amsfonts}%
\usepackage{amsthm}%
\usepackage{mathrsfs}%
\usepackage[title]{appendix}%
\usepackage{xcolor}%
\usepackage{textcomp}%
\usepackage{manyfoot}%
\usepackage{booktabs}%
\usepackage{algorithm}%
\usepackage{algorithmicx}%
\usepackage{algpseudocode}%
\usepackage{listings}%

\usepackage{subfig}
\usepackage[percent]{overpic}

\usepackage{booktabs} % for \toprule \midrule \bottomrule
\usepackage{makecell} % for multiple rows in a table
\usepackage{multirow}
\usepackage{multicol}
\usepackage{threeparttable} % for adding notes in a table
\usepackage{lscape}

\usepackage{bm} % bold symbol
\usepackage{bbm}  % lowercase \mathbb{}
\usepackage[thicklines]{cancel} % cancel line

\usepackage{adjustbox}
\usepackage{float}
\usepackage{soul}

\usepackage{chngcntr}
\usepackage{enumitem}

\usepackage[mathlines]{lineno}
\usepackage{etoolbox}
\pretocmd{\equation}{\linenomathNonumbers}{}{}
\apptocmd{\endequation}{\endlinenomath}{}{}
\pretocmd{\align}{\linenomathNonumbers}{}{}
\apptocmd{\endalign}{\endlinenomath}{}{}
\pretocmd{\gather}{\linenomathNonumbers}{}{}
\apptocmd{\endgather}{\endlinenomath}{}{}

\theoremstyle{thmstyleone}%
\theoremstyle{thmstyletwo}%

\theoremstyle{thmstylethree}%

\makeatletter
\hypersetup{hypertexnames=false}
\makeatother

\begin{document}

\title[Article Title]{Mechanistic multiphysics modeling reveals how blood pulsation drives CSF flow, pressure, and brain deformation under physiological and injection conditions}

% Mechanistic_multiphysics_modeling_reveals_how_blood_pulsation drives_CSF_flow_pressure_and_brain_deformation_under_physiological_and_injection_conditions


\author[1]{\fnm{Zhuogen} \sur{Li}}

\author[1]{\fnm{Keyu} \sur{Feng}}

\author*[1,2]{\fnm{Hector} \sur{Gomez}}\email{ hectorgomez@purdue.edu}

\affil[1]{\orgdiv{School of Mechanical Engineering}, \orgname{Purdue University}, \orgaddress{\street{585 Purdue Mall}, \city{West Lafayette}, \postcode{47906}, \state{Indiana}, \country{USA}}}

\affil[2]{\orgdiv{Weldon School of Biomedical Engineering}, \orgname{Purdue University}, \orgaddress{\street{206 S
Martin Jischke Dr}, \city{West Lafayette}, \postcode{47906}, \state{Indiana}, \country{USA}}}


\abstract{
Intrathecal (IT) injection is an effective way to deliver drugs to the brain bypassing the blood-brain barrier. To evaluate and optimize IT drug delivery, it is necessary to understand the cerebrospinal fluid (CSF) dynamics in the central nervous system (CNS). In combination with experimental measurements, computational modeling plays an important role in reconstructing CSF flow in the CNS. Existing models have provided valuable insights into the CSF dynamics; however, most neglect the effects of tissue mechanics, focus on partial geometries, or rely on measured CSF flow rates under specific conditions, leaving full-CNS CSF flow field predictions across different physiological states underexplored. Here, we propose a comprehensive multiphysics computational model of the CNS with three key features: (1) it is implemented on a fully closed geometry of CNS; (2) it includes the interaction between CSF and poroelastic tissue as well as the compliant spinal dura mater; (3) it has potential for predictive simulations because it only needs data on cardiac blood pulsation into the brain.
Our simulations under physiological conditions demonstrate that our model accurately reconstructs the CSF pulsation and captures both the craniocaudal attenuation and phase shift of CSF flow along the spinal subarachnoid space (SAS). When applied to the simulation of IT drug delivery, our model successfully captures the intracranial pressure (ICP) elevation during injection and subsequent recovery after injections.
The proposed multiphysics model provides a unified and extensible framework that allows parametric studies of CSF flow dynamics and optimization of IT injections, serving as a strong foundation for integration of additional physiological mechanisms.
}




\maketitle

% \linenumbers

\section{Introduction}\label{sec:introduction}

As the processing center of the human body, the central nervous system (CNS), consisting of the brain and spinal cord, integrates sensory input and issues commands to coordinate humans' behavior and maintain homeostasis, through intricate networks of neurons and glial cells \cite{Kandel2013-CNS-function}. The treatment of CNS disorders, such as dementia, Parkinson’s disease, and stroke, has long been a major focus in medicine because damage to the brain or spine often leads to severe disability or even death, while the current therapies are still limited \cite{WHO2006-neurological}. As one of the treatment approaches, intrathecal (IT) injections can bypass the blood-brain barrier (BBB), which restricts the entry of most therapeutic agents into the CNS, and deliver drugs through the surrounding cerebrospinal fluid (CSF) to the target region \cite{Tangen2018-IT-review}. During and after IT injections, the drug disperses within CSF, which primarily flows in the subarachnoid space (SAS), a narrow and elongated region between the arachnoid mater and the pia mater. However, the drug transport in SAS may be impeded by periodic CSF oscillations with low net flow and reduced by non-target absorption or clearance via CSF circulation \cite{Mazur2019-IT-review, De-Andres2022-IT-review}. These phenomena necessitate strategies to optimize the IT injections; in this context, it is essential to understand the dynamics of CSF, which serves as the primary driver of drug transport. CSF velocity, flow patterns, and pathological variations critically affect the therapeutic performance. In addition, analyzing the pressure propagation in the CNS may help investigate the cause of IT injections' side effects. Because the brain is an extremely sensitive organ, understanding how IT injections affect intracranial pressure (ICP) will help de-risk and scale IT injections.

In vivo animal experiments (non-invasive magnetic resonance imaging (MRI) \cite{Rivera-Rivera2024-MRI-review} or invasive measurements \cite{Liu2022-in-vivo}) and in vitro models \cite{Tangen2017-in-vitro, Khani2020-in-vitro} have proven effective to study CSF dynamics, but each approach has inherent limitations. Although MRI is non-invasive and provides high spatial resolution, it cannot directly measure the pressure evolution or monitor the transient state during IT injections. Invasive in vivo animal experiments preserve physiological conditions but face interspecies differences and lack parameter control. In vitro models allow parameter control and reproducibility but oversimplify the complex CNS anatomy. Moreover, all these experimental approaches require high equipment and labor costs. This underscores the complementary value of computational methods because they are noninvasive, inexpensive, highly replicable, and easily parameterizable, allowing exploration of a wide range of physiological and pathological conditions. References \cite{Linninger2016-review, Kurtcuoglu2019-review} provide comprehensive reviews of computational models of the CNS.

Existing computational approaches can be evaluated from three perspectives: physics framework, anatomical coverage, and modeling approach. From the perspective of the physics framework, computational fluid dynamics (CFD) is predominant since there are mature CFD solvers. Early CFD simulations modeled CSF flow in idealized geometries \cite{Kurtcuoglu2005-idealized-ventricles, Linge2014-idealized} of CSF-filled spaces to explore basic flow patterns. With the development of high-resolution MRI, researchers have been able to adopt patient-specific geometries \cite{Causemann2022-fsi-brain, Sass2017-spine-geometry}, enabling more accurate predictions of CSF flow and pressure distributions. Recently, researchers have incorporated microanatomical structures \cite{Stockman2006-fine-structures, Gupta2009-flow-resistance, Heidari-Pahlavian2014-fine-structures, Tangen2015-microanatomy}, such as arachnoid trabeculae, nerve roots, and denticulate ligaments, into the computational geometries to investigate their effects on CSF flow patterns. In spite of these advances, most CFD models impose rigid-wall boundary conditions on the CNS tissues, neglecting the effects of tissue compliance. The drawbacks of the rigid-wall assumptions are listed in \cite{Heidari-Pahlavian2015-drawbacks}. This simplification may fail to capture two important phenomena in the spinal SAS: amplitude attenuation and phase shift of CSF flow waveforms. To address the drawbacks of CFD modeling, researchers conduct fluid-structure interaction (FSI) simulations, coupling CSF with deformable CNS tissues. Solid mechanics models of CNS tissues include poroelasticity \cite{Causemann2022-fsi-brain, Linninger2009-fsi-poroelasticity} or poro-viscoelasticity \cite{Gholampour2018-fsi} of brain tissue, linear elasticity \cite{Cheng2014-fsi} of spinal cord, and hyperelasticity \cite{Sweetman2011-hyperelastivity-dura} of the spinal dura mater. In terms of anatomical coverage, many CFD and FSI simulations are restricted to partial anatomies on truncated CNS segments, such as individual cranial \cite{Causemann2022-fsi-brain} and spinal subsystems \cite{Cheng2014-fsi}, superior \cite{Gupta2010-superior-cSAS} or inferior \cite{Gupta2009-flow-resistance} cranial SAS, the ventricular system \cite{Howden2008-ventricular-system}, and sections of the spinal SAS \cite{Roldan2009-cervical-SAS, Yiallourou2012-cervical-SAS, Rutkowska2012-cervical-SAS, Helgeland2014-cervical-SAS}. Partial models allow high-resolution simulations due to their small computational cost. However, such subsystems need artificial boundary conditions at the truncations (CSF velocity and/or traction), which are not easily measurable. Moreover, drug transport simulations on partial models are problematic because the boundary conditions of drug concentration at the truncations are unable to capture the drug exchange with the neglected region. Therefore, full CNS models \cite{Sweetman2011-hyperelastivity-dura, Howden2011-full-CNS, Tangen2015-microanatomy} that reconstruct the entire CSF-filled spaces are necessary for studying global CSF flow and simulating drug transport. With respect to the modeling approach, we differentiate between {\it mechanism-oriented} models and {\it results-oriented} models. A mechanism-oriented model is one that, based on prior evidence, postulates the underlying mechanisms of CSF dynamics, and then compares the model results with measurements that are different from those used to establish the prior evidence. A mechanism-oriented model is ideal to understand the underlying physical and biological mechanisms and predict outcomes from first principles. A results-oriented model is one that aims at accurately reproducing certain measurements, often combining mechanistic elements with data-driven tuning to match a target outcome. While results-oriented models are undoubtedly useful in practical applications, they prioritize agreement with experimental data over mechanistic generality, and they may be unreliable when used to interrogate questions different from those used to develop the model. We categorize most existing CFD models as results-oriented. One reason is that they account for the compliance of the spinal dura mater by using pre-defined moving boundary conditions, rather than computing the displacement of the dura mater via simulation. 
For example, reference \cite{Kuttler2010-moving-bc} imposed a moving boundary condition on the dura mater to match the steady transport velocity in the spinal SAS from experiments. {While the agreement with the experiment is good, the ability of this model to generalize to other situations remains unclear.} In addition, \cite{Khani2017-moving-BC} measured CSF flow rates at several levels of the spinal SAS and prescribed radial displacements of the dura mater matching the difference in flow rate between two adjacent levels. This approach allows the simulated CSF flow rate to closely match experimental data, but it relies on the assumption that the dura mater's displacement is uniform between two adjacent levels, which is only valid when these two levels are close to each other. Thus, densely measured flow rate data is needed. Moreover, such approaches confine the simulation to the physiological condition where the experimental data is acquired. Any change in the physiological condition, such as IT injections, alters the CSF flow pattern and makes the original measurements inapplicable. 

The three evaluations above show that CFD, partial, or results-oriented models have limitations if used as mechanistic, predictive models of CSF dynamics and highlight the need for alternative approaches. In this study, we address this gap and develop a mechanistic, predictive multiphysics model of the CNS. The model operates on a closed full CNS geometry and accounts for fluid flow, tissue deformations and their coupled interactions. Specifically, we couple CSF flow with the poroelasticity of the CNS tissue and the elasticity of the spinal dura mater. We also include the mechanical effects of microanatomical structures in the SAS and elastic support from the epidural fat. The CSF pulsation in our model is driven by pulsatile blood flow into the brain, which only requires the measurements of blood flow rate. This enables generalization to different conditions, including IT injections, for which we also incorporate CSF absorption to better capture the CSF flow and pressure changes.

We show that our mechanism-oriented model, based on first principles only,  captures the amplitude attenuation and phase shift of CSF flow waveforms in spinal SAS that occur in physiological conditions. We also use our model to simulate the effect of IT injections on the CNS response. Our results show that the model reproduces the ICP rise that occurs during injection and its subsequent decay, which is driven by CSF reabsorption. The unique features of the model offer an opportunity to quantitatively predict ICP during IT injection, which is critical to de-risk and scale this administration modality. In all, we believe that our multiphysics model offers a powerful platform for studies of CSF flow mechanisms, enabling investigations of different pathological conditions and injection strategies. Furthermore, our model allows extensions that incorporate additional physiological driving mechanisms.

\section{Multiphysics model of the CNS}

\subsection{From physiological mechanisms to computational modeling}\label{sec:physiological-to-computational}

\subsubsection{Physiology of CSF dynamics}\label{sec:physiological-background}

\begin{figure}[htb!]
    \centering
    \includegraphics[width=\linewidth]{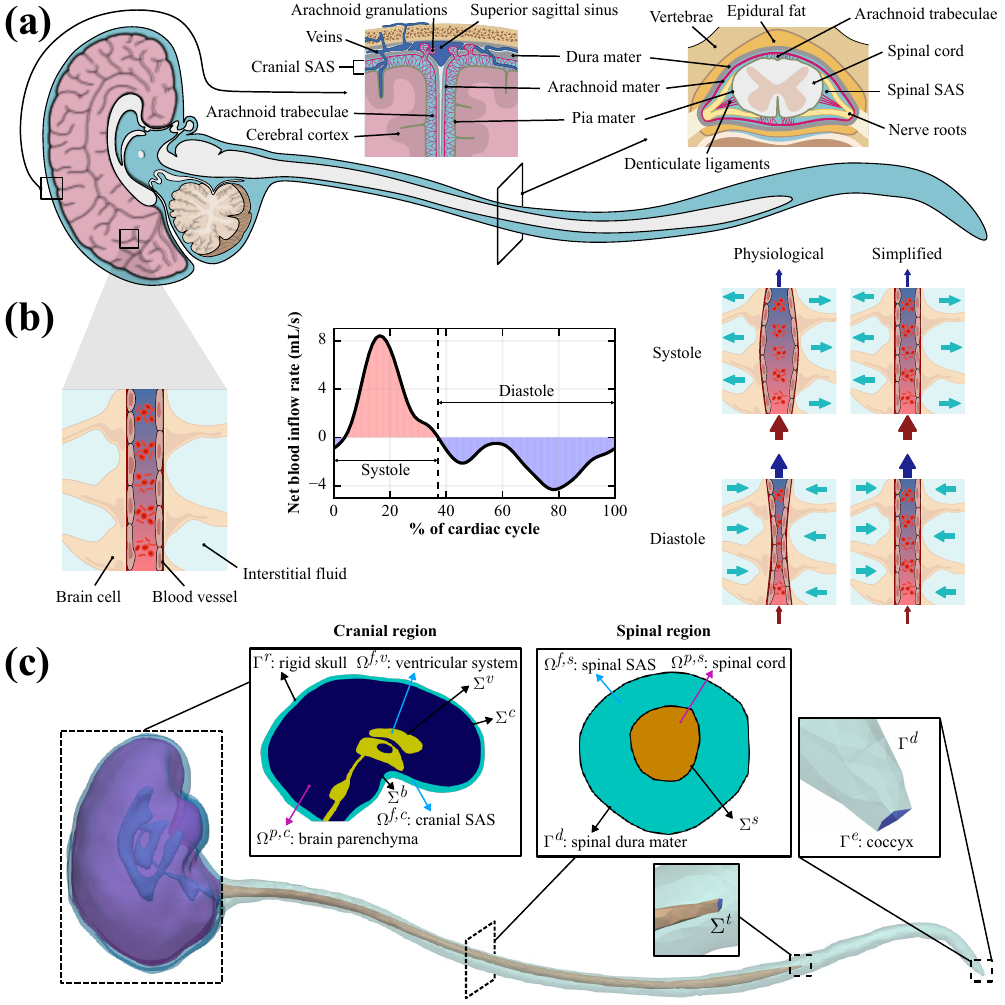}
    \caption{(a) Illustration of the central nervous system. (b) \textit{Left}: Blood-brain interface. Blood flows in vessels surrounded by porous brain tissue filled with interstitial fluid. \textit{Middle}: Net blood inflow rate into the brain fitted from \cite{Baledent2014-blood-inflow}. \textit{Right}: Simplification of the driver of the CSF pulsation. The expansion during systole and contraction during diastole of blood vessels is simplified as ISF production during systole and absorption during diastole, respectively. (c) Computational domain used in this study. The domain can be partitioned anatomically or into different subdomains of materials: CSF-filled spaces ($\Omega^{f}$), including ventricular system ($\Omega^{f,v}$), cranial SAS ($\Omega^{f,c}$), and spinal SAS ($\Omega^{f,s}$); porous tissue ($\Omega^{p}$), including brain parenchyma ($\Omega^{p,c}$) and spinal cord ($\Omega^{p,s}$); an elastic membrane representing the spinal dura mater ($\Gamma^d$), the caudal end of which is marked by $\Gamma^e$.}
    \label{fig:comp-domain}
\end{figure}

Fig.~\ref{fig:comp-domain}(a) illustrates a sagittal view of the CNS with two insets: one for the lateral view of the cranial SAS and the other for the axial view of the spinal SAS. The CSF flows pulsatilely in the ventricular system and subarachnoid spaces, mainly due to the pulsation of the cardiac blood flow in the brain \cite{Haughton2014-cardiac-driven, Daouk2017-cardiac-driven}. 
According to the Monro-Kellie doctrine \cite{Cushing1926-MK-doctrine}, the intracranial bulk volume of nervous tissue, blood and fluid remains constant because the skull is rigid. Therefore, the periodic change in intracranial blood volume during the cardiac cycle is primarily compensated by pulsatile displacement of CSF toward the spinal SAS, where the compliance is considerably higher than in the cranium. 
The higher compliance of the spinal compartment arises from the deformability of the spinal dura mater, which is enabled by the soft tissue and veins in the epidural space \cite{Linninger2016-review}. 
In addition to cardiac pulsation, respiratory activity and body posture also influence the CSF pulsations. 
The local CSF flow in the SAS and the tissue deformation are further affected by the presence of microstructures such as cranial arachnoid trabeculae, spinal nerve roots and denticulate ligaments, which form a connective network between the arachnoid and pia maters. The cranial arachnoid trabeculae are slender collagenous filaments that anchor the arachnoid mater to the pia mater---they introduce flow resistance to CSF motion and elastic support to the brain parenchyma. The spinal nerve roots and denticulate ligaments connect the spinal cord to the dura mater, providing elastic support and limiting their displacement within the spinal SAS.
The magnitude of the CSF pulsation is of the order of 1 mL/s \cite{Greitz2004-CSF, Bhadelia1997-CSF-pulsation}. Simultaneously, CSF is continuously produced by the choroid plexus in the ventricular system and reabsorbed into venous circulation mainly through the arachnoid granulations at the upper cranial SAS \cite{Linninger2016-review}. Production and reabsorption occur at rates below 0.01 mL/s \cite{Greitz2004-CSF, Johanson2008-CSF-production}, over two orders of magnitude smaller than the pulsatile rate.
As a result, CSF exhibits a bidirectional pulsatile motion along the craniospinal axis and serves to balance volume changes between the cranial and spinal compartments. These physiological mechanisms serve as a basis for building a reliable multiphysics model of the CNS and predicting the CSF flow dynamics accurately.

\subsubsection{Modeling framework}\label{sec:modeling-framwwork}

To capture the pulsatile CSF flow dynamics, we idealize the CNS as a closed system driven only by the net blood inflow into the cranium, defined as the difference between the arterial inflow and the venous outflow. The respiratory and postural effects are neglected. 
We also neglect the continuous CSF production and reabsorption under physiological conditions, since their rates are much smaller than the CSF pulsation rate. 
The details of the net blood inflow rate that we adopt, denoted by \( \dot{Q}^{\rm{blood}} \), are shown in Fig.~\ref{fig:comp-domain}(b) and Appendix \ref{secA-net-blood-inflow}.

The periodic net volumetric blood inflow rate represents the instantaneous change in the intracranial blood volume that induces CSF motion according to Monro-Kellie doctrine. To avoid explicitly modeling the complex vascular network of the brain, we formulate the net blood inflow as an equivalent interstitial fluid (ISF) production and absorption within the brain parenchyma \cite{Causemann2022-fsi-brain}. This formulation preserves the mechanical effect of blood inflow on parenchymal expansion and fluid pressure elevation. In addition, this equivalent ISF source is assumed to be spatially uniform throughout the brain parenchyma, which simplifies the computation while maintaining cranial volume conservation; see Eq.~\eqref{eq:porous-source-g}.

The driving mechanism above is incorporated into our multiphysics model for the CNS that couples the CSF/ISF flow with the deformation of the CNS tissues. The CSF-filled spaces are regarded as a fluid domain (Sec.~\ref{sec:GovEqn-CSF}), while the brain parenchyma and the spinal cord are represented as deformable poroelastic tissue saturated with ISF (Sec.~\ref{sec:GovEqn-poroelastic}). The CSF domain connects to the brain parenchyma and the spinal cord through the pia mater and the ventricular walls. Here, we neglect the pia mater as a separated layer. The outer boundary of our model is defined by the arachnoid-dura bilayer, which is hereafter referred to collectively as the dura mater for simplicity. The cranial dura mater is assumed to be a rigid wall because it is supported by the rigid skull, while the spinal dura mater is modeled as a two-dimensional elastic membrane (Sec.~\ref{sec:GovEqn-dura}).

Although necessary for stabilizing the CSF pathways and neural tissues, the fine anatomical structures in the SAS, including cranial arachnoid trabeculae, spinal nerve roots and denticulate ligaments, are not geometrically resolved; instead, we incorporate their mechanical effects into our model. The cranial arachnoid trabeculae are represented as a flow resistance acting on CSF (Eq.~\eqref{eq:flow-resistance}) and a one-dimensional elastic support to the brain parenchyma (Eq.~\eqref{eq:winkler}). The spinal nerve roots and denticulate ligaments are modeled together as a one-dimensional elastic support to the spinal cord (Eq.~\eqref{eq:winkler}). In the epidural space, we assume that there is only epidural fat and model it as a one-dimensional elastic support to the spinal dura mater (Sec.~\ref{sec:GovEqn-dura}).

In addition to the physiological conditions, we model intrathecal injections as a volumetric source added to the CSF domain. To accurately predict the pressure increase induced by the injections, we introduce a volumetric sink to represent the pressure-driven CSF absorption through cranial arachnoid granulations. The detailed formulations of intrathecal injections and CSF absorption are presented in Sec.~\ref{sec:IT-injection}.

\subsection{Computational domains and boundaries}

We merge the brain geometry from \cite{Causemann2022-fsi-brain} and the spine geometry from \cite{Sass2017-spine-geometry} to construct a full closed CNS geometry, which serves as our computational domain \( \Omega \); see Fig.~\ref{fig:comp-domain}d. 
The computational domain \( \Omega \) consists of three subdomains, where we solve for different physics: (1) the CSF-filled spaces (\( \Omega^{f} \)), including the ventricular system (\( \Omega^{f,v} \)), cranial SAS (\( \Omega^{f,c} \)), and spinal SAS (\( \Omega^{f,s} \)); (2) the porous tissue filled with ISF (\( \Omega^{p} \)), containing the brain parenchyma (\( \Omega^{p,b} \)) and spinal cord (\( \Omega^{p,s} \)); and (3) the spinal dura mater (\( \Gamma^{d} \)), which is regarded as a 2D elastic membrane surrounding the spinal SAS. 
We follow \cite{Putluru2025-dura-thickness} and truncate the spinal cord geometry below the conus medullaris, as its size there is negligible.

The outer boundary of the spinal SAS coincides with the spinal dura mater (\( \Gamma^{d} \)), due to its 2D formulation. The outer boundary of the cranial SAS, representing the cranial dura mater, is denoted by \( \Gamma^r \). We mark the caudal end of the spinal dura mater with \( \Gamma^e \), which attaches to the coccyx. 
The interface between the CSF and poroelastic tissue is denoted by \( \Sigma \), including five non-overlapping components: (1) \( \Sigma^v \), connecting the ventricular system and brain parenchyma; (2) \( \Sigma^c \), connecting the SAS and brain parenchyma excluding the brain stem; (3) \( \Sigma^b \), connecting the SAS and brain stem; (4) \( \Sigma^s \), representing the lateral interface between the SAS and spinal cord together with the brain stem; and (5) \( \Sigma^t\), denoting the caudal end surface of the spinal cord where it is truncated.
The unit outward normal vectors to the boundaries of \( \Omega^{f} \) and \( \Omega^{p} \), and to \( \Gamma^{d} \) are denoted by \( \bm{n}^f \), \( \bm{n}^p \), and \( \bm{n}^d \), respectively. 
It follows from the topological arrangement of the subdomains that \( \bm{n}^f = - \bm{n}^p \) on \( \Sigma \) and \( \bm{n}^p = \bm{n}^d \) on \( \Gamma^{d} \).

\subsection{Governing equations}

\subsubsection{Cerebrospinal fluid flow}\label{sec:GovEqn-CSF}

We model CSF as an incompressible Newtonian fluid. We neglect the nonlinear advection effect due to the low Reynolds number in CSF flow \cite{Causemann2022-fsi-brain}. To account for the flow resistance produced by the arachnoid trabeculae in the cranial SAS without explicitly resolving their geometry, we model the cranial SAS as a Brinkman porous medium \cite{Brinkman1949-porous}, which imposes a Darcy-type drag on the CSF flow \cite{Gupta2009-flow-resistance}. Therefore, the governing equations of the CSF are given by the unsteady Stokes-Brinkman formulation
\begin{subequations}\label{eq:CSF}
\begin{alignat}{2}
   \rho^f \frac{\partial \bm{u}^f}{\partial t}
   -\nabla \cdot \bm{\sigma}^f
   + \bm{S}^{f}
   & = \bm{0}, &\quad& \text{in } \Omega^f \times [0,T],
   \label{eq:CSF-COLM} \\
   \nabla \cdot \bm{u}^f  & = s, &\quad& \text{in } \Omega^f \times [0,T]. \label{eq:CSF-COM} 
\end{alignat}
\end{subequations}
Here, \( \rho^f \) is the CSF density, \( \bm{u}^f \) is the CSF velocity and \( \bm{\sigma}^f \) is the Cauchy stress tensor 
\begin{align}\label{eq:cauchy-stress-csf}
    \bm{\sigma}^f = - p^f \bm{I} + \mu^f \left[ \nabla \bm{u}^f + \left( \nabla \bm{u}^f \right)^T \right].
\end{align}
In Eq.~\eqref{eq:cauchy-stress-csf}, \( p^f \) is the CSF pressure,  \( \bm{I} \) is the identity tensor, and \( \mu^f \) is the CSF dynamic viscosity.
The source term \(s\) in Eq.~\eqref{eq:CSF-COM} represents volumetric fluid addition or removal. We assume $s=0$ under physiological conditions. However, when we model intrathecal injections, $s$ takes a nonzero value determined by the injection characteristics and the CSF absorption rate; see Sec.~\ref{sec:IT-injection}.

In Eq.~\eqref{eq:CSF-COLM}, \( \bm{S}^{f} \) denotes the flow resistance induced by the arachnoid trabeculae in the cranial SAS. We assume that the flow resistance is transversely-isotropic---the resistance is isotropic in the two trabecular-transverse axes but differs from that along the trabecular-longitudinal axis. In the ventricular system and the spinal SAS, we take \( \bm{S}^f = \bm{0} \). Under these assumptions, the flow resistance term is given by
\begin{align}\label{eq:flow-resistance}
    \bm{S}^{f} =
    \begin{aligned}
    \begin{cases}
        \mu^f \left[ \dfrac{1}{\kappa^l} \left( \bm{n}^c \otimes \bm{n}^c \right) + \dfrac{1}{\kappa^t} \left( \bm{I} - \bm{n}^c \otimes \bm{n}^c \right) \right] \cdot \bm{u}^f , &\text{in } \Omega^{f,c} \times [0,T], \\
         \bm{0}, &\text{otherwise},
    \end{cases}
    \end{aligned}
\end{align}
where \( \bm{n}^c \) denotes the unit vector along the longitudinal axis of a trabecula; see Appendix \ref{secA-FEM-AT}. The parameters \( \kappa^l \) and \( \kappa^t \) represent, respectively, the longitudinal and transverse permeability. They are derived analytically from the phase-averaged form of Navier-Stokes equations over a representative unit cell which preserves porous properties and captures details of trabecular structures  \cite{Westhuizen1994-flow-resistance, Gupta2009-flow-resistance}. They can be expressed as
\begin{align}
    \kappa^l = \frac{r^2 (1-V_t)^2 ( \pi + 2.157 V_t)}{ 48 V_t^2},
    \quad
    \kappa^t = \frac{\pi r^2 (1-V_t) ( 1 - \sqrt{V_t} )^2}{ 24 V_t^{3/2}},
\end{align}
where \( r \) is the average radius of the trabeculae and \( V_t \) is the volume fraction of trabecular tissue in the cranial SAS.

\subsubsection{Brain parenchyma and spinal cord poromechanics}\label{sec:GovEqn-poroelastic}

We model the porous tissue, including the brain parenchyma and spinal cord, as linear poroelastic media fully saturated with ISF. The coupling between fluid flow and deformation in linear poroelastic materials is described by the Biot equations \cite{Biot1941-Biot-theory}
\begin{subequations}\label{eq:porous}
\begin{alignat}{2}
    - \nabla \cdot \bm{\sigma}^p & = 0, &\quad& \text{in } \Omega^p \times [0,T],
    \label{eq:porous-COLM} \\
    c\frac{\partial p^p}{\partial t} 
    + \alpha \frac{\partial (\nabla \cdot \bm{d}^p)}{\partial t} 
    - \nabla \cdot \left( \frac{\kappa}{\mu^f} \nabla p^p  \right)
    & = g, &\quad& \text{in } \Omega^p \times [0,T]. \label{eq:porous-COM}
\end{alignat}
\end{subequations}
Here, \( c \) is the specific storage coefficient, \( p^p \) is the pore ISF pressure, \( \alpha \) is the Biot-Willis coefficient, \( \bm{d}^p \) is the tissue displacement, and \( \kappa \) is the permeability. The Cauchy stress tensor \( \bm{\sigma}^p \) takes the form
\begin{align}
    \bm{\sigma}^p = \lambda^p ( \nabla \cdot \bm{d}^p ) \bm{I} 
    + \mu^p \left[ \nabla \bm{d}^p + \left( \nabla \bm{d}^p \right)^T \right]
    - \alpha p^{p} \bm{I},
\end{align}
where \( \lambda^p \) and \( \mu^p \) are the first and second Lam\'e parameters of the solid skeleton. In Eq.~\eqref{eq:porous-COM}, \( g \) is the equivalent spatially-uniform ISF production rate in the brain parenchyma \cite{Causemann2022-fsi-brain} and is given by
\begin{align}\label{eq:porous-source-g}
    g = 
    \begin{aligned}
    \begin{cases}
         \dfrac{\dot{Q}^{\text{blood}}}{V^p}, &\text{in } \Omega^{p,b} \times [0,T], \\
         0, &\text{otherwise},
    \end{cases}
    \end{aligned}
\end{align}
where \( \dot{Q}^{\text{blood}} \) denotes the net blood inflow rate into the brain and \( V^p = \int_{\Omega^{p,b}}\mathrm{d}V \) represents the volume of the brain parenchyma; see Appendix \ref{secA-net-blood-inflow} for the exact expression of \( \dot{Q}^{\text{blood}} \) used in our computations.

\subsubsection{Spinal dura mater elasticity}\label{sec:GovEqn-dura}

Because the thickness of the spinal dura mater is much smaller than its other dimensions, we model the spinal dura as a 2D linear elastic membrane. We assume that the dura mater is under plane stress augmented by the transverse shear stress, neglecting any variations across its thickness \cite{Figueroa2006-cmm}. This condition indicates that the out-of-plane normal stress and all the normal derivatives are zero. 
To implement this condition, we define the normal and tangential projection tensor
\begin{align}
    \bm{P}^n = \bm{n}^d \otimes \bm{n}^d,
    \quad
    \bm{P}^{\tau} = \bm{I} - \bm{P}^n = \bm{I} - \bm{n}^d \otimes \bm{n}^d,
\end{align}
where the supersripts \(n\) and \(\tau\) denote normal and tangential, respectively. 
We define the in-plane del operator \(\nabla_{\tau} = \bm{P}^{\tau} \cdot \nabla \) to remove the normal component from the standard del operator.
The surface load applied to the dura mater is treated as a volumetric load uniformly distributed along its thickness, due to the thin membrane assumption.
The linear elasticity of the dura mater is finally described as
\begin{align}
    \rho^d \frac{\partial \bm{u}^d}{\partial t}
    - \nabla_{\tau} \cdot \bm{\sigma}^d
    = \frac{1}{\xi} \left(
    \bm{t}^f + \bm{t}^{\text{pre}} + \bm{t}^e \right).
    \label{eq:dura-COLM}
\end{align}
Here, \(\rho^d\) is the density of the dura mater, \( \bm{u}^d \) is its velocity, and \( \xi \) is its thickness. The Cauchy stress tensor \( \bm{\sigma}^d \) takes the form
\begin{align*}
    \bm{\sigma}^d 
    = 
    \lambda^d ( \nabla_{\tau} \cdot \bm{d}^d) \bm{I}
    +
    \mu^d \left[ \bm{J}^{\text{plane}} +(\bm{J}^{\text{plane}})^T \right]
    + k^d \mu^d \left[ \bm{J}^{\text{shear}} +(\bm{J}^{\text{shear}})^T \right],
\end{align*}
where \( \bm{J}^{\text{plane}} = \bm{P}^{\tau} \,\nabla_{\tau} \bm{d}^d \,\bm{P}^{\tau} \) and \( \bm{J}^{\text{shear}} = \bm{P}^{\tau} \,\nabla_{\tau} \bm{d}^d \,\bm{P}^n + \bm{P}^n \,\nabla_{\tau} \bm{d}^d \,\bm{P}^{\tau} \). The constants \( \lambda^d \) and \( \mu^d \) are the first and second Lam\'e parameters under the plane stress condition, \( k^d \) is the shear correction factor, accounting for the variation of the transverse stress along the normal direction of the membrane \cite{Figueroa2006-cmm, Hughes2000-fembook}. 
In Eq.~\eqref{eq:dura-COLM}, \( \bm{t}^f \) is the CSF traction, \( \bm{t}^{\text{pre}} \) is the pre-traction balancing the initial CSF pressure load (see Sec.~\ref{sec:GovEqn-ICBC}), and \( \bm{t}^{\text{e}} = -k^e (\bm{n}^d \otimes \bm{n}^d) \cdot \bm{d}^d   \) represents the elastic support from the epidural fat \cite{Moireau2012-ext-tiss-sup}, where \( k^e \) is its elastic constant. Note that we only account for the normal support from the epidural fat and neglect its damping effect.

\subsubsection{Interface and transmission conditions}\label{sec:GovEqn-interface}

At the interface between the CSF and the spinal dura mater, which coincides with the latter, we adopt the following conditions
\begin{subequations}\label{eq:interface}
\begin{alignat}{2}
  \bm{u}^f &= \bm{u}^d,    &\quad& \text{on }\Gamma^d \times [0,T], \label{eq:interface-dof}  \\
  \bm{\sigma}^f\cdot\bm{n}^f + \bm{t}^f &= \bm{0}, 
  &\quad & \text{on }\Gamma^d \times [0,T]. \label{eq:interface-stress}
\end{alignat}
\end{subequations}
Eq.~\eqref{eq:interface-dof} enforces kinematic compatibility between the CSF and spinal dura, and \eqref{eq:interface-stress} enforces the traction balance.

At the interface between the CSF and the porous tissue, we adopt the following transmission conditions \cite{Ruiz-Baier2022-eye, Causemann2022-fsi-brain}
\begin{subequations}\label{eq:transmission}
\begin{alignat}{2}
    \bm{u}^f \cdot \bm{n}^f + 
    \left(
    \frac{\partial \bm{d}^d}{\partial t}
    -
    \frac{\kappa}{\mu^f} \nabla p^{p}
    \right) \cdot \bm{n}^p
    & = 0, &\quad& \text{ on } \Sigma \times [0,T], 
    \label{eq:transmission-flux} \\
    \bm{\sigma}^f \cdot \bm{n}^f + \bm{\sigma}^p \cdot \bm{n}^p - \bm{f}^{W} &= \bm{0}, &\quad& \text{ on } \Sigma \times [0,T], 
    \label{eq:transmission-stress} \\
    \bm{n}^f \cdot \bm{\sigma}^f \cdot \bm{n}^f + p^{p} &= 0, &\quad& \text{ on } \Sigma \times [0,T], 
    \label{eq:transmission-normal}\\
    (\bm{I} - \bm{n}^f \otimes \bm{n}^f) \cdot
    \left[
    \bm{\sigma}^f \cdot \bm{n}^f 
    +
    \frac{\gamma \mu^f}{\sqrt{\kappa}}
    \left(
    \bm{u}^f - \frac{\partial \bm{d}^p}{\partial t}
    \right)
    \right]
    &= \bm{0}, &\quad& \text{ on } \Sigma \times [0,T]. 
    \label{eq:transmission-tangential} 
\end{alignat}
\end{subequations}
Here, \eqref{eq:transmission-flux} denotes the fluid mass balance across the interface, \eqref{eq:transmission-stress} accounts for the balance of the momentum, \eqref{eq:transmission-normal} represents the balance of the fluid normal stress, and \eqref{eq:transmission-tangential} stands for the Beavers-Joseph-Saffman condition \cite{Beavers1967-BJF, Saffman1971-BJF} for the tangential fluid traction, where \( \gamma \) is the slip rate coefficient. In Eq.~\eqref{eq:transmission-stress}, \( \bm{f}^W \) is the traction from the small structures in the SAS and reads
\begin{align}\label{eq:winkler}
    \bm{f}^W = 
    \begin{cases}
        -k^c \bm{d}^p, &\text{in } \Sigma^{c} \times [0,T], \\
        -k^c (\bm{n}^p\otimes\bm{n}^p) \cdot \bm{d}^p, &\text{in } \Sigma^{b} \times [0,T], \\
        -k^s (\bm{n}^p\otimes\bm{n}^p) \cdot \bm{d}^p, &\text{in } \Sigma^{s} \times [0,T], \\
         0, &\text{otherwise},
    \end{cases}
\end{align}
where \( k^c \) and \( k^s \) are the elastic coefficient of the arachnoid trabeculae in the cranial SAS and the microstructures in spinal SAS, respectively.

\subsubsection{Initial and boundary conditions}\label{sec:GovEqn-ICBC}

We assume that the system is initially at rest \cite{Causemann2022-fsi-brain} with 
\begin{subequations}
\begin{alignat}{3}
    \bm{u}^f &= 0, &\quad p^f &= p_0^f, &\quad& \text{in } \Omega^f \times \{ 0 \},
    \\
    \bm{d}^p &= 0, &\quad p^p &= p_0^f, &\quad& \text{in } \Omega^p \times \{ 0 \},
\end{alignat}
\end{subequations}
where \( p_0^f \) denotes the initial pressure of the CSF and pore ISF, which we assume to be the same. The velocity and displacement initial conditions are required to start the computations. However, they do not influence the simulation results because we only analyze the results after the system becomes periodic. %In principle, these initial conditions can alternatively be removed because they only serve to initialize the numerical solver, rather than represent any physiological state.

The pre-traction balancing the initial CSF pressure on the spinal dura mater in Sec.~\ref{sec:GovEqn-dura} is given by
\begin{align}
    \bm{t}^{\text{pre}} = -p_0^f \bm{n}^d.
\end{align}

Due to the rigid support from the skull and the coccyx, we assume no-slip boundary conditions at the cranial dura mater \( \Gamma^r \) and the caudal end of the spinal dura mater \( \Gamma^e \), that is,
\begin{align}
    \bm{u}^f = 0, \quad \text{on } (\Gamma^r \cup \Gamma^e) \times [0,T].
\end{align}

\subsubsection{Intrathecal injection and compensatory CSF absorption}\label{sec:IT-injection}
During intrathecal injections, the needle is inserted through the skin, the subcutaneous tissue, and the dura mater to deliver the drug into the SAS. Resolving the needle geometry and the fine-scale jet from the needle tip requires a very fine mesh around the needle, which leads to high computational cost. Because the injection rate is low for patients' safety, we can assume that the needle jet only perturbs the local flow field with negligible effects on distant locations. Therefore, we define an injection subdomain \( \Omega^{f,\text{injt}} \) whose characteristic length exceeds the needle diameter, and we represent the injection via a volumetric source term \(   s^{\text{injt}} \) within \( \Omega^{f,\text{injt}} \), which is given by
\begin{align}\label{eq:source-injection}
    s^{\text{injt}} = \frac{\dot{Q}^{\text{injt}}}{V^{\text{injt}}}.
\end{align}
Here, \( \dot{Q}^{\text{injt}} \) denotes the volumetric injection rate and \( V^{\text{injt}} = \int_{\Omega^{f,\text{injt}}}\;\mathrm{d}\Omega \) represents the volume of the injection subdomain.

During the injection, the ICP increases due to the added fluid volume and the limited compliance of the CNS. It subsequently decreases as the CSF is reabsorbed into the venous circulation. The CSF reabsorption occurs mainly through the arachnoid granulations (AG) \cite{Linninger2016-review}, which are located along the upper cranial SAS. This reabsorption process is unidirectional \cite{Eklund2007-RC,Proulx2021-CSF-absorption} and pressure-driven, with the absorption rate approximately proportional to the pressure difference between the CSF and the venous blood \cite{Albeck1991-absorption-coeff,Ekstedt1978-absorption-coeff}.
In our model, we treat the fluid mass change due to CSF reabsorption similarly to the mass change produced by injections. We assume that CSF reabsorption occurs only at AGs and define a reabsorption subdomain \( \Omega^{f,\text{absp}} \) in the upper cranial SAS with a characteristic length larger than that of the AGs. We represent the absorption via a volumetric sink term to avoid fine-scale meshing around the AGs and the associated high computational cost. 
Furthermore, we set this sink term as a pressure-dependent one-way valve to mimic the physiological unidirectional CSF absorption mechanism in the upper cranial SAS, allowing flow only when the CSF pressure exceeds the valve threshold and preventing backflow.
The sink term \(\dot{Q}^{\text{absp}}\) is given by
\begin{align}\label{eq:absorption-term}
   \dot{Q}^{\text{absp}} = \max \{\beta ( p^f - p^{f,\text{th}} ), 0 \},
\end{align}
where \( p^{f,\text{th}} \) is the threshold pressure which equals the maximum CSF pressure under physiological conditions, and \( \beta \) is a positive absorption coefficient obtained from experimental data; see Table~\ref{tab:parameters}. 

The volumetric sink term of absorption is given by
\begin{align}\label{eq:sink-absorption}
    s^{\text{absp}} = \frac{\dot{Q}^{\text{absp}}}{V^{\text{absp}}},
\end{align}
where \( \dot{Q}^{\text{absp}} \) denotes the volumetric absorption rate and \( V^{\text{absp}}  = \int_{\Omega^{f,\text{absp}}}\;\mathrm{d}\Omega \) represents the volume of the absorption subdomain. 

Assuming that the drug solution has the same material properties as CSF, the source term \(s\) in \eqref{eq:CSF-COM} is formulated based on Eqs. \eqref{eq:source-injection} and \eqref{eq:sink-absorption} as
\begin{align}
    s = 
    \begin{cases}
    s^{\text{injt}}, & \text{in } \Omega^{f,\text{injt}}, \\
    s^{\text{absp}},  & \text{in } \Omega^{f,\text{absp}}, \\
    0,     & \text{otherwise}.
    \end{cases}
\end{align}

\subsection{Model parameters}
The model parameters used in this study, as well as reference values from literature, are listed in Table~\ref{tab:parameters}.

\begin{landscape}
\begin{table}[h]
\centering
\caption{Parameters used in this study.}
\label{tab:parameters}
\begin{tabular}{@{}llllll@{}}
\toprule
\textbf{Parts} & \textbf{Symbol}  & \textbf{Description} &  \textbf{Value} & \textbf{Unit} & \textbf{Ref. value} \\
\midrule
CSF & \( \rho^f \)  &  Density  & 1007  & kg/m\(^3\) & 1007 \cite{Barber1970-density} \\
    & \( \mu^f \)   &  Dynamic viscosity  & \(0.8\times10^{-3}\) & \(\text{Pa}\cdot\text{s}\) & \(0.7\times10^{-3}\) - \(10^{-3}\)\cite{Bloomfield1998-viscosity}\\
\midrule
Porous tissue & \( E^p \)  & Young's modulus 
              & \makecell[tl]{1500 (BP\footnotemark[1]) \\ 3000 (BS\footnotemark[1], SC\footnotemark[1])} 
              & Pa & 1895 (WM\footnotemark[1]); 1389 (GM\footnotemark[1]) \cite{Budday2015-Young-modulus}\\
              & \( \nu^p \)  & Poisson's ratio  & 0.479  & - & 0.479 \cite{Smith2007-brain-tissue} \\
              & \( \lambda^p \) & 1st Lam\'e parameter & Calculated\footnotemark[2]
              & Pa &  \\
              & \( \mu^p \) & 2nd Lam\'e parameter & Calculated\footnotemark[2]
              & Pa &  \\
              & \( \alpha \) & Biot-Willis coefficient & 1.0 & - & 1.0 \cite{Smith2007-brain-tissue} \\
              & \( c \) & Storage coefficient & \( 1.0\times10^{-6} \) & \( \text{Pa}^{-1} \) & \makecell[tl]{\(4.47\times10^{-7}\) \cite{Chou2016-storage-coeff} \\ \(1.5\times10^{-5}\) - \(3\times10^{-4}\) \cite{Guo2019-storage-coeff} } \\
              & \( \kappa \) & Permeability & 100 & \( \text{nm}^{2} \) & 10 - 4000 \cite{Holter2017-permeability} \\
              & \( \gamma \) & Slip-rate coefficient & 1.0 & - & 0.01 - 5 \cite{Ehrhardt2012-slip-rate-coeff}\\
\midrule
Spinal dura mater  & \( \rho^d \)  & Density & 1174 & kg/m\(^3\)
                   & 1174 \cite{Mcintosh2010-tissue-param} \\
                   & \( E^d \)  & Young's modulus\footnotemark[3] & 2.0 & MPa
                   & 4 - 144 \cite{Runza1999-dura-elasticity}\\
                   & \( \nu^d \)  & Poisson's ratio & 0.5 & -
                   &  \\
                   & \( \lambda^d \) & 1st Lam\'e parameter & Calculated\footnotemark[4]
                   & Pa &  \\
                   & \( \mu^d \) & 2nd Lam\'e parameter & Calculated\footnotemark[4]
                   & Pa &  \\
                   & \( \xi \)  & Thickness & 0.2 & mm
                   & 0.2 \cite{Cavelier2022-dura-properties} \\
                   & \( k^d \)  & Shear correction factor  & 5/6  & - & 5/6 \cite{Hughes2000-fembook} \\
\midrule
Cranial arachnoid trabeculae  & \( r \)  & Radius & 15 & \(\mu\)m
                            & 15.75 \cite{Benko2020-AT-distribution}\\
                            & \( V_t \)  & Volumetric fraction  & 0.25 & - & 0.2556 \cite{Benko2020-AT-distribution} \\
                            & \( k^c \) & Elasticity coefficient  & 12.6 & kPa/mm &  12.6 \cite{Benko2021-trabeculae-modulus} \\
\midrule
Spinal microstructures  & \( k^s \)  & Elasticity coefficient  & \(1.0\times10^{6}\) & Pa/m & \\
\midrule
Epidural fat & \( E^e \)  & Young's modulus & 1.6 & kPa  
             &  1.6 \(\pm\) 0.8 \cite{Alkhouli2013-fat-modulus}  \\
             & \( l^e \)  & Thickness       & 4 & mm
             &  4 \cite{Putluru2025-dura-thickness} \\
             & \( k^e \)  & Elastic coefficient  & \(E^e/l^e\) & kPa/mm &  \\
\midrule
Arachnoid granulations & \( \beta \)  & CSF absorption coefficient  & \(1.5\times10^{-11}\) & m\(^{3}\)/\((\text{s}\cdot\text{Pa})\) &   \(1.5\times10^{-11}\) \cite{Albeck1991-absorption-coeff} \\
\midrule
Initial condition  & \( p^{f}_{0} \)  & Pressure at rest & 600 & Pa & 667 - 2000 \cite{Rangel-Castilla2008-pressure-range}\\
\botrule
\end{tabular}
\footnotetext[1]{BP = brain parenchyma; BS = brain stem; SC = spinal cord; WM = white matter; GM = gray matter.}
\footnotetext[2]{\( \lambda^p = E^p\nu^p/\left[(1+\nu^p)(1-2\nu^p)\right]\); \( \mu^p = E^p/\left[2(1+\nu^p)\right] \).}
\footnotetext[3]{The value reported in \cite{Runza1999-dura-elasticity} is a tangent elastic modulus at large strain that is not representative of the Young's modulus. The data in \cite{Runza1999-dura-elasticity} shows that the Young's modulus (not reported) is smaller than the tangent elastic modulus at large strain that they report.}
\footnotetext[4]{\( \lambda^d = E^d\nu^d/\left[(1+\nu^p)(1-\nu^d)\right]\); \( \mu^d = E^d/\left[2(1+\nu^d)\right] \).}
\end{table}
\end{landscape}

\subsection{Computational method}
Our multiphysics model is discretized in time using the backward Euler scheme. We discretize in space with the finite element method (FEM), following \cite{Ruiz-Baier2022-eye, Causemann2022-fsi-brain}. 
The total number of elements is 419,906, with 134,682 in the fluid domain and 285,224 in the poroelastic domain. The model is solved with the open-source finite element package \textit{FeniCS} \cite{AlnaesEtal2015-fenics} with its multiphysics extension \textit{multiphenics} \cite{multiphenics}. The details of the solution are elaborated in Appendix \ref{secA-FEM}. 

\section{Results}
\subsection{CSF dynamics under physiological conditions}\label{sec:results-physiological}
We run simulations under physiological conditions (without intrathecal injections) for 20 cardiac cycles. After 20 cardiac cycles the system reaches periodic behavior within a relative error of 1\%. Results are shown for the final cycle. We focus on the capability of our model to predict the CSF flow and tissue movement in the spinal SAS, where the tissue compliance plays a critical role. The compliance determines how much the spinal dura mater can deform in response to the CSF pressure oscillations and regulates the amplitude and phase of CSF flow along the spinal SAS. Thus, accurate modeling of the tissue compliance is essential for capturing the correct propagation of CSF flow, which cannot be achieved in rigid-wall simulations.

\subsubsection{CSF flow exhibits craniocaudal decay and phase shift in the spinal SAS}
To quantify the pulsatile CSF flow dynamics in the spinal SAS, we compute the CSF flow rate across different spinal levels as
\begin{align}
\dot{Q}^{f}_{s}=\int_{\Gamma^{f}_{s}}\bm{}\bm{u}^f\cdot\bm{n}^{f}_{s}\mathrm{\;d}A,
\end{align}
where \(\Gamma^{f}_{s}\) denotes the cross sections at different levels of spinal SAS with the unit normal caudocranial vector \(\bm{n}^{f}_{s}\). Here, we assume that the change in \(\Gamma^{f}_{s}\) is negligible in flow rate calculation and thus we define \(\Gamma^{f}_{s}\) in the undeformed configuration of the spinal SAS. The time integral of \(\dot{Q}^{f}_{s}\) within one cardiac cycle is called the cumulative CSF flow.

The spinal CSF flow rate represents the redistribution of the intracranial pulsation along the spinal SAS. The craniocaudal changes in its amplitude and phase reflect the attenuation and delay of the pulsation transport affected by the spinal compliance. Therefore, the quantification of these variations directly provides insight into the mechanical response of the spinal compliance to the CSF pulsation.

\begin{figure}[htbp]
    \centering
    \includegraphics[width=\linewidth]{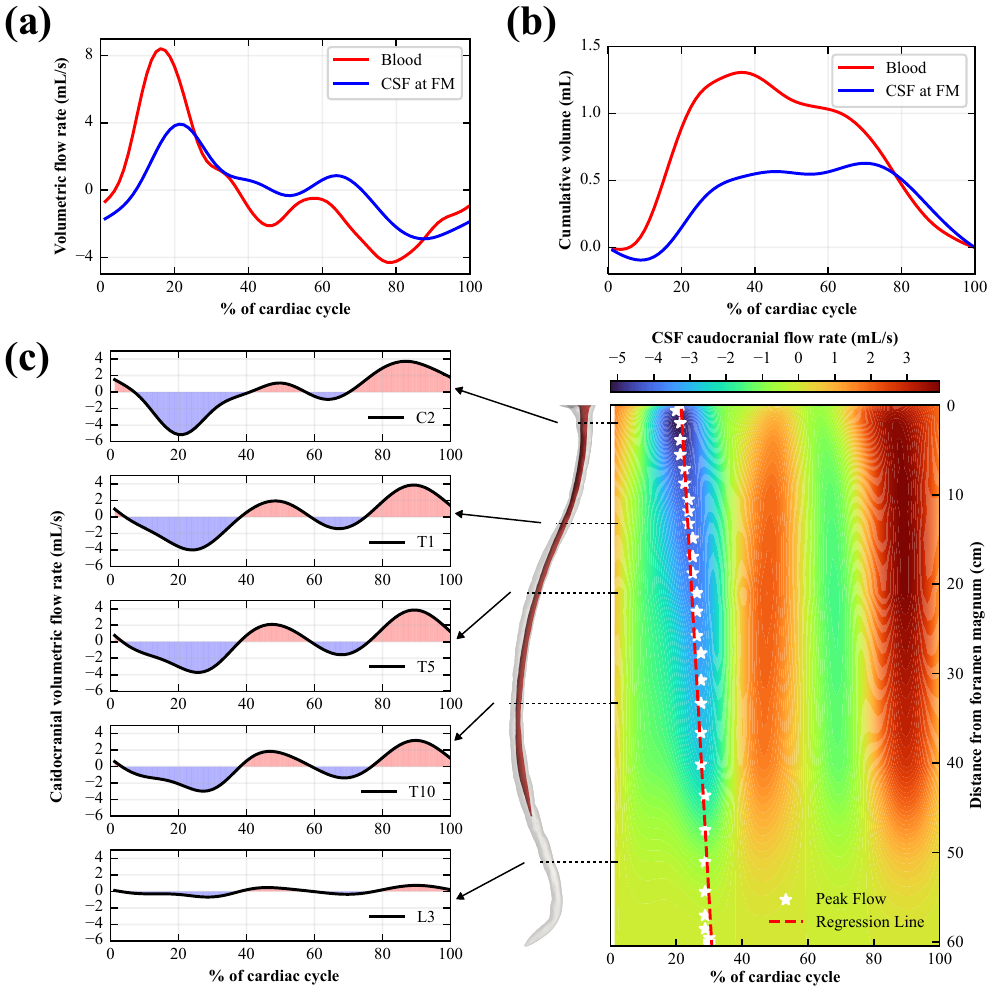}
    \caption{(a) Instant volumetric flow rate of net blood inflow and CSF craniocaual flow across the foramen magnum. (b) Cumulative blood inflow into the brain and craniocaual flow across the foramen magnum. (c) \textit{Left}: CSF caudocranial flow rate at C2, T1, T5, T10, and L3 level. \textit{Right}: CSF caudocranial waveforms along the spinal SAS. We mark the time instants of the peak craniocaudal flow with white stars and fit them with a linear curve. The estimated slope of the curve, representing the pulse wave speed, is 6.47 m/s.}  
    \label{fig:flow-rate}
\end{figure}

To analyze the CSF flow pattern, we first compare the net blood inflow into the brain with the CSF flow across the foramen magnum (FM). 
Figures \ref{fig:flow-rate}(a) and (b) illustrate their instantaneous volumetric flow rates and the corresponding cumulative volumes in a cardiac cycle. They reveal an amplitude decay and a phase lag of the flow rate of CSF across the foramen magnum relative to the net blood inflow. In our model, these two features result from the combined effects of the fluid storage in the brain parenchyma that is quantified by the storage coefficient \(c\), and the additional flow resistance momentum \(\bm{S}^f\) in Eq.~\eqref{eq:CSF-COLM}. The storage coefficient \(c\) measures the volumetric compliance of the brain. According to the simulation in \cite{Causemann2022-fsi-brain}, a larger storage coefficient causes a significant decrease in the CSF flow rate across the foramen magnum. In addition, \(\bm{S}^f\) serves as a resistance to the CSF flow, which not only dampens the flow rate magnitude, but also produces the phase lag of CSF flow.

We now focus on the CSF flow in the spinal SAS. Fig.~\ref{fig:flow-rate}(c) shows the spatio-temporal distribution of CSF caudocranial flow rate in one cardiac cycle. The white stars denote the peak craniocaudal flow rate. We also give the CSF flow rate waveforms at C2, T1, T5, T10, and L3 levels. Along the spinal SAS, our model captures the craniocaudal decay and phase shift of the flow rate, which are observed in vivo \cite{Sass2017-spine-geometry, Coenen2019-spinal-SAS-experiment}. Following \cite{Sass2017-spine-geometry}, we fit a linear curve of the time instants of peak craniocaudal flow rate, the slope of which gives the pulse wave speed (PWS) approximately equal to 6.47 m/s, lying within the experimentally measured range (1.19 - 9.19 m/s) summarized in \cite{Putluru2025-dura-thickness}. The PWS characterizes the propagation speed of the CSF pulsation through the compliant spinal canal and thus serves as an indicator of the overall compliance of the CSF system.

\subsubsection{CSF pressure shows craniocaudal amplification and flow-coupled phase shift in the spinal SAS}
To characterize the spatiotemporal behavior of the CSF pressure in the spinal SAS,
we compute the cross-sectional average CSF pressure
\begin{align}
p^{f}_{s}=\frac{\int_{\Gamma^{f}_{s}}\bm{}p^f\mathrm{\;d}A}{\int_{\Gamma^{f}_{s}}\mathrm{d}A},
\end{align}
at different spinal levels. The distribution of this average pressure affects the CSF flow and reflects how the spinal compliance responds to the CSF pulsation.

\begin{figure}[htb!]
    \centering
    \includegraphics[width=\linewidth]{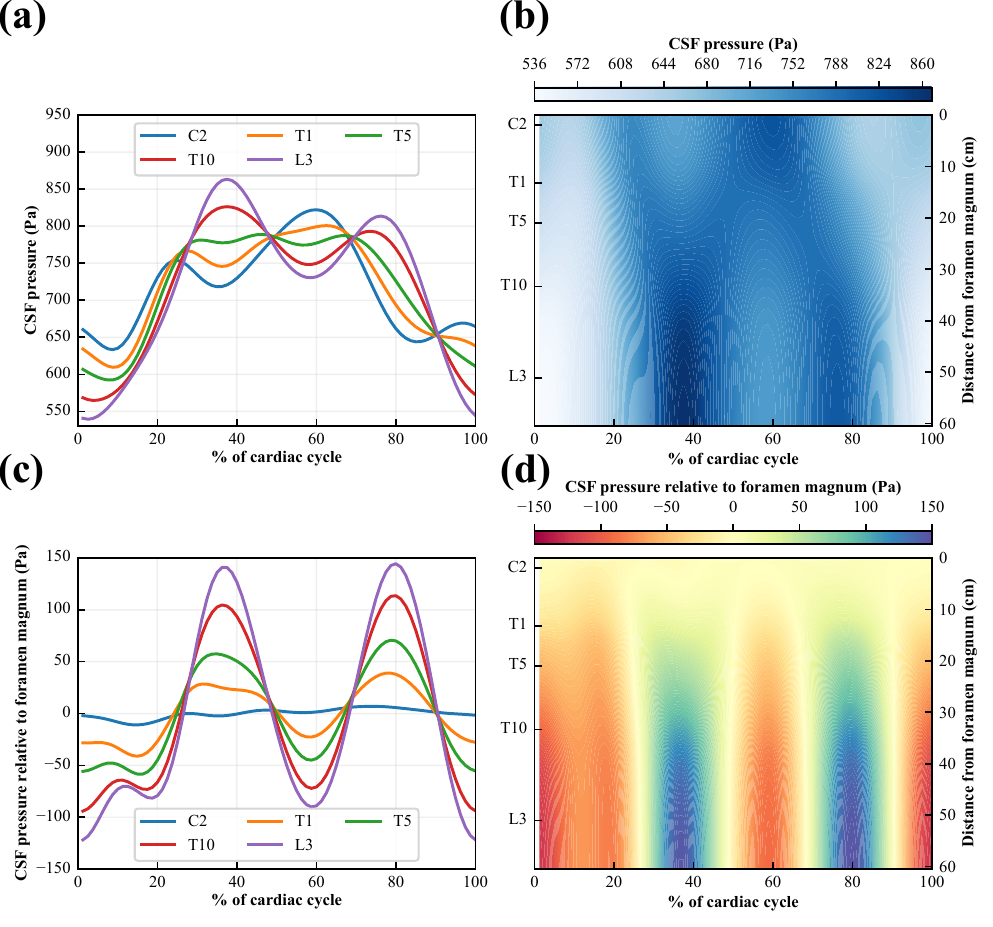}
    \caption{(a) CSF pressure at C2, T1, T5, T10, L3 levels in one cardiac cycle. (b) CSF pressure profile in spinal SAS in one cardiac cycle. (c) CSF pressure relative to foramen magnum at C2, T1, T5, T10, L3 levels in one cardiac cycle. (d) CSF pressure relative to foramen magnum in spinal SAS in one cardiac cycle.}
    \label{fig:pressure_spine}
\end{figure}

Fig.~\ref{fig:pressure_spine}(a) and (b) show a cyclic CSF pressure oscillation in the spinal SAS, sharing the same period as the cardiac cycle. The CSF pressure amplitude increases from the foramen magnum toward the caudal end, indicating a craniocaudal amplification of the pressure wave. Fig.~\ref{fig:pressure_spine}(c) and (d) show the instantaneous pressure difference relative to the foramen magnum, which varies approximately within \([-150\text{ Pa}, 150\text{ Pa}]\) and increases craniocaudally along the spinal SAS. Figures \ref{fig:flow-rate} and \ref{fig:pressure_spine} show a temporal correlation between the CSF flow and pressure and reveal a flow-coupled phase shift: when the CSF moves caudally, the CSF pressure decreases from cranial to caudal levels, and vice versa. At flow reversal points, the pressure gradient reaches a local maximum, while near the flow peaks, the pressure becomes nearly uniform along the spinal axis.

\subsubsection{Spinal tissue displacements attenuate craniocaudally}

The motions of the spinal cord and the spinal dura mater represent the response of the spinal tissue to the CSF pulsation and enable the CNS to redistribute and buffer the CSF flow in the spinal SAS. Therefore, it is essential to understand the spinal tissue displacements for explaining the craniocaudal decay of CSF flow and validating the ability of our model to capture physiologically realistic CNS dynamics.

\begin{figure}[htb!]
    \centering
    \includegraphics[width=\linewidth]{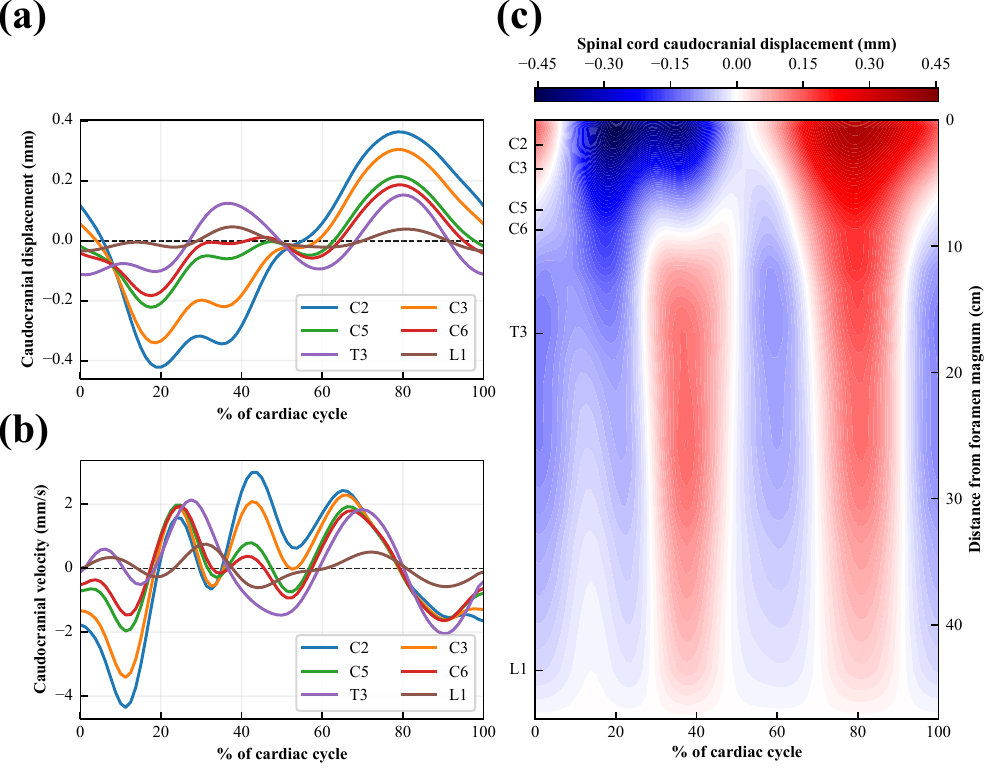}
    \caption{(a) Caudocranial displacement of spinal cord at C2, C3, C5, C6, T5, L1 levels. (b) Caudocranial velocity of the spinal cord at C2, C3, C5, C6, T5, L1 levels. (c) Caudocranial displacement of the spinal cord in spinal SAS. The region below the bottom axis is neglected because the spinal cord is truncated there in our model.} 
    \label{fig:tissue-displacement}
\end{figure}

We first focus on the spinal cord and specifically analyze its caudocranial displacements; see Fig.~\ref{fig:tissue-displacement}. From Fig.~\ref{fig:tissue-displacement}(a) and (c) we see that at the cervical region, the spinal cord is moving craniocaudally and then caudocranially in one cardiac cycle, following the CSF flow pattern. The peak craniocaudal displacement is approximately 0.45 mm and the magnitude of the displacement decreases craniocaudally. At the thoracic and lumbar regions, the spinal cord has four movement reversal points. Fig.~\ref{fig:tissue-displacement}(b) shows the velocity of the spinal cord at different spinal levels. 
{The simulated peak-to-peak velocity magnitude at C2 (7.36 mm/s) and C3 (5.70 mm/s) agree well with the in vivo measurement at the C2/C3 region ($5.80\pm2.0$ mm/s)} in \cite{Wolf2023-sc-displacement}. However, {the simulated values at C5 (3.94 mm/s) and C6 (3.58 mm/s) are lower than the experimental measurement at the C5/C6 region ($6.78\pm2.8$ mm/s)}. 
This may be because the spinal cord motion is also affected by the respiration or cardiac-induced momentum, which is neglected in our simulation.

\begin{figure}[htb!]
    \centering
    \begin{overpic}[width=\linewidth]{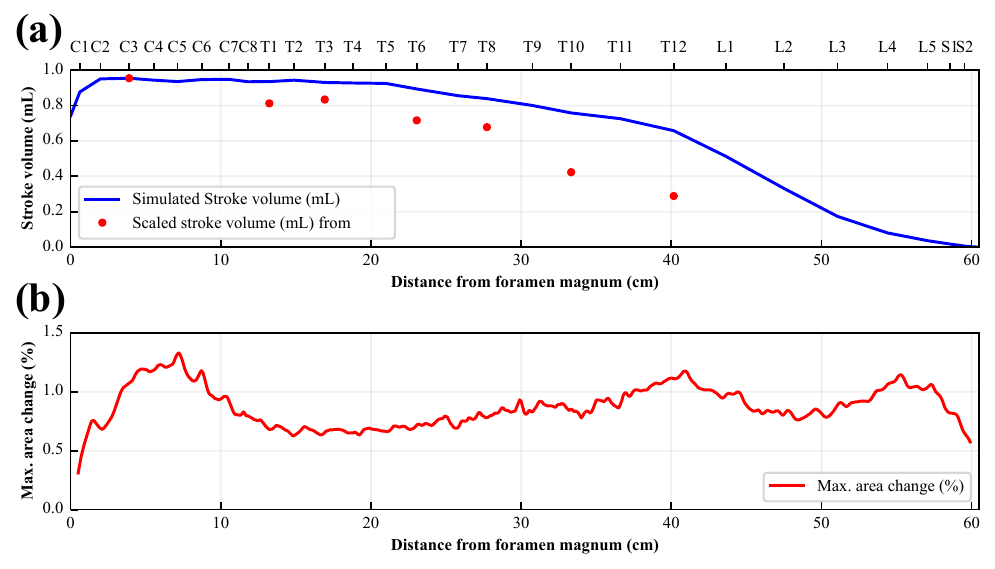}
        \put(35.5,34.15){\footnotesize \cite{Coenen2019-spinal-SAS-experiment}}
    \end{overpic}
    \caption{(a) Comparison of stroke volume along the spinal SAS between our simulations and MRI measurements \cite{Coenen2019-spinal-SAS-experiment}. The in vivo measured stroke volumes, obtained at C3, T1, T3, T6, T8, T10 and T12 levels, are scaled so that the value at C3 matches the simulated one, facilitating comparison of spatial trends along the spinal SAS. (b) Maximum percentage area change of different levels of cross-section in the spinal SAS.}
    \label{fig:Quantities}
\end{figure}

We now look at the motion of the spinal dura. We are interested in two quantities: (1) stroke volume; and (2) maximum percentage change in the cross sectional area of the spinal SAS. The stroke volume is defined as half of the integral of the magnitude of the CSF flow rate across a spinal level over a cardiac cycle and is given by
\begin{align}
\text{Stroke volume}
= 
\frac{1}{2}\int_{T_{\text{cc}}}\left\vert\dot{Q}_{s}^{f}\right\vert\mathrm{\;d}t,
\end{align}
where \(T_{\text{cc}}\) denotes a full cardiac cycle. The stroke volume represents the net oscillatory fluid volume exchange across different spinal levels and serves as a measure of the compliance of the spine. Fig.~\ref{fig:Quantities}(a) shows the stroke volume in the spinal SAS. The stroke volume slightly increases from the foramen magnum to the cervical region. This is because of two factors. First, the flow resistance produced by the trabeculae in the cranium attenuates the CSF flow rate. Second, the CSF flows from the stiff cranium to the compliant spine, which enhances the aggregation of CSF at the entrance of the cervical region. At the thoracic and lumbar regions, the stroke volume decreases craniocaudally, with a faster attenuation in the lumbar region. {We also compare our simulated stroke volume with the MRI-measured stroke volume} \cite{Coenen2019-spinal-SAS-experiment}. {Because the stroke volume is patient-specific, we cannot compare the values directly; instead, we scale the MRI-measured stroke volume so that its value at C3 matches the simulation, which allows the comparison of their trend along the spinal SAS. The simulated stroke volume exhibits a craniocaudal decreasing trend similar to that of the MRI measurements, highlighting the importance of spinal compliance in CSF flow prediction. However, the rate of decrease is slightly higher in the MRI measurements. This discrepancy can be attributed to the nonuniform craniocaudal distribution of the epidural fat along the spinal SAS} \cite{DeAndres2011-epidural-fat-distribution}, which is thicker in the lower spinal region than in the upper region. Thicker epidural fat provides weaker elastic support to the spinal dura mater and increases the local spinal compliance. Since our model assumes a uniform epidural fat distribution, the spinal compliance at the thoracic region is lower in our simulation than in reality, which accommodates less CSF motion and explains the slower decrease in the predicted thoracic stroke volume. 

The maximum percentage change in the cross-sectional area of the spinal SAS is defined as
\begin{align}
    \%\Delta A_s = \frac{\left|\int_{\Gamma^{f}_{s}(t_{\%}=t_{\%}^\star)}\mathrm{d}A-\int_{\Gamma^{f}_{s}(t_{\%}=0)}\mathrm{d}A\right|}{\int_{\Gamma^{f}_{s}(t_{\%}=0)}\mathrm{d}A},
\end{align}
where \(t_{\%}\in[0,100]\) denotes the time percentage of a cardiac cycle, and \(t_{\%}^\star\) is the time percentage that maximizes the quantity $\int_{\Gamma^{f}_{s}(t_{\%})}\mathrm{d}A$. Fig.~\ref{fig:Quantities}(b) shows the maximum percentage area change in spinal SAS. We can see that at the cervical and lumbar regions, the percentage area change is larger (approximately 1\%) than at the thoracic region (approximately 0.5 - 1\%). 

\subsection{CSF dynamics during and after intrathecal injections}\label{sec:results-injection}

\subsubsection{Case study}
\begin{figure}[htb!]
    \centering
    \includegraphics[width=\linewidth]{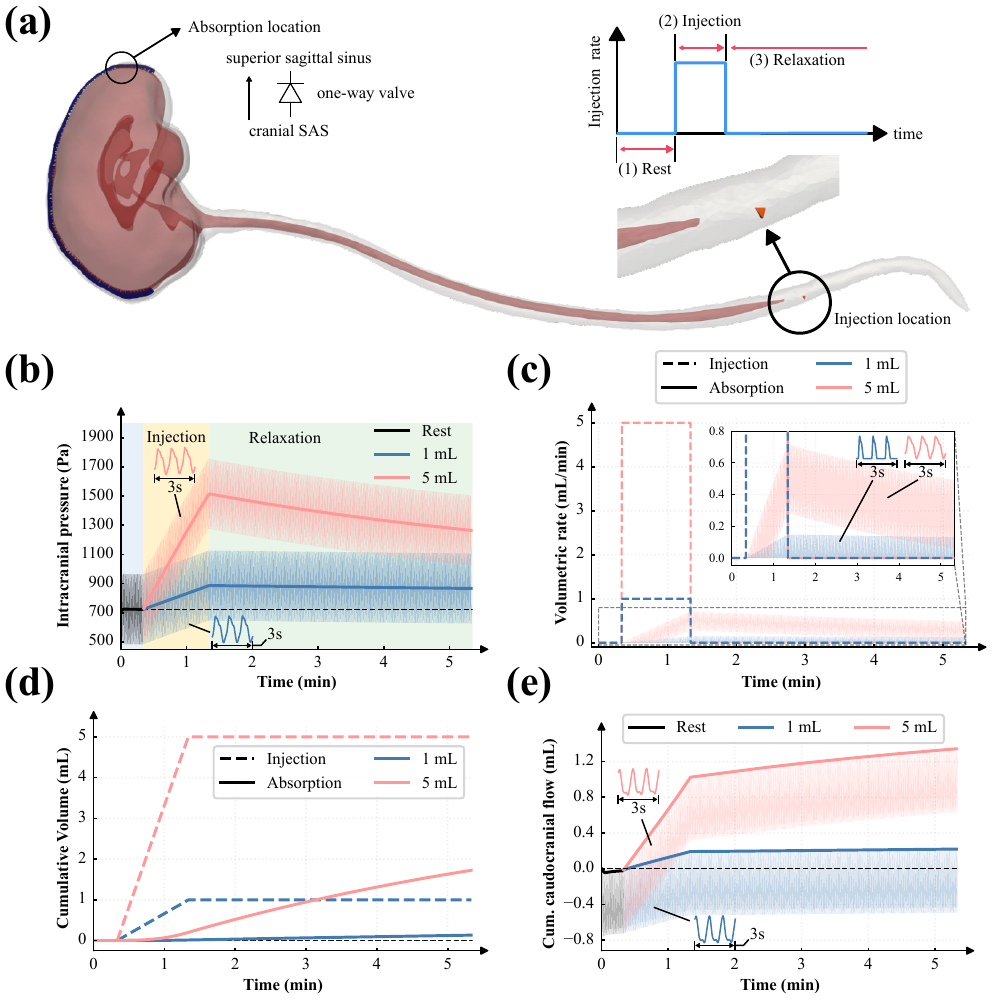}
    \caption{(a) Illustration of the simulation for intrathecal injections. We set the injection location at L2 level and apply a constant injection rate during the injections. At the arachnoid granulations, we set the absorption as a one-way valve. The quantities of interest are shown in (b)-(e). Here, we consider two cases of 1-minute injections: 1 mL (blue) and 5 mL (red). The shaded regions in (b), (c), and (e) represent the cardiac-induced oscillating quantities. (b) ICP profile before, during, and after the injections. The averages of the upper and lower envolopes, marked by solid curves, represent the ICP elevation and recovery. The 1 mL and 5 mL injections raise ICP by 161.4 Pa and 784.5 Pa, respectively. (c) Injection rate (dashed curves) and absorption rate (solid curves). (d) Cumulative injection volume and absorption volume. (e) Cumulative volumetric caudocranial flow across the foramen magnum. The upper envolopes, shown as solid curves, denote a bulk caudocranial movement of CSF due to the CSF absorption in the cranial region.}
    \label{fig:injection}
\end{figure}

IT injections are usually administered at the lower lumbar region to protect the spinal cord. In this study, we set the injection location at L2 level and set the absorption region that represents the arachnoid granulations at the upper cranial SAS; see Fig.~\ref{fig:injection}(a). We start the simulation of the injections at the end of the simulation for physiological conditions, considering two cases of constant-rate injection: (1) 1 mL in 1 min; (2) 5 mL in 1 min. Then, we continue each simulation for an additional 4 min to investigate the relaxation phase after the injections. 

We monitor the change of the ICP, the CSF volumetric flux across the foramen magnum, and the volume of CSF absorption in the arachnoid granulations.

\subsubsection{Intracranial pressure elevates with injection and recovers via CSF absorption}
The ICP reflects the balance between brain tissue, cerebral blood, and CSF. An abnormal ICP increase or decrease will cause cerebral circulatory disorders \cite{Raboel2012-review-ICP-measurement} and potentially brain damage. Currently, most effective methods for ICP monitoring are invasive \cite{Raboel2012-review-ICP-measurement, Rajajee2024-noninvasive}. Therefore, a computational prediction of ICP change is a very valuable tool for de-resking IT injections. Here, we quantify the effect of the injections on ICP by monitoring the ICP change during and after the injections.

We define ICP as
\begin{align}
    \text{ICP} =  \frac{1}{V^p}\int_{\Omega^{p,c}} p^p\;\mathrm{d} \Omega,
\end{align}
which is equal to the volumetric average pore pressure in the brain parenchyma. Fig.~\ref{fig:injection}(b) shows the time evolution of the ICP before, during, and after the injection. Because the ICP oscillates with the cardiac cycle frequency, the plots at a time scale of $\sim\,$5 min look like a shaded region, but the insets in the figure (time window of $\sim\,$3 s) reveal the true time evolution of the ICP. The thicker lines denote the average of the upper and lower envelopes of the ICP curves. Fig.~\ref{fig:injection}(c) shows the injection rate in the lumbar region and the absorption rate in the arachnoid granulations.
Fig.~\ref{fig:injection}(d) shows the corresponding cumulative injection and absorption volume.

During the injections, the injection rate exceeds the absorption rate, increasing the total fluid volume in the CSF-filled spaces. Thus, the average ICP increases due to the compliance of the spine. Meanwhile, the absorption rate increases as the ICP increases, according to Eq.~\eqref{eq:absorption-term}. When the injection ends, the ICP reaches its peak. The 1 mL injection raises ICP by 161.4 Pa, while the 5 mL injection produces a 784.5 Pa increase. The ratio \(784.5 \text{ Pa}/161.4\text{ Pa}\approx4.84\) is slightly smaller than \(5\text{ mL}/1\text{ mL}=5\), which suggests a slight sublinear ICP increase with respect to the injection rate. 
After the injections, the ICP decreases nearly exponentially, supporting the idea that the system behaves like a first-order RC circuit \cite{Eklund2007-RC}. Specifically, the CSF absorption results in ICP decrease, which successively decreases the CSF absorption rate. Thus, the decreasing CSF absorption rate causes the exponential trend of ICP decrease. The trends of ICP elevation and recovery agree well with the low-rate infusion experiments for rats in \cite{Belov2021-experiments-injection}.

\subsubsection{Caudocranial CSF flow across the foramen magnum increases during injection due to CSF absorption}

To quantify how the injections and the compensatory absorptions affect the bulk CSF flow, we calculate the cumulative caudocranial flow across FM; see Fig.~\ref{fig:injection}(e). Since the cardiac cycle starts near the beginning of the systole, the CSF flows craniocaudally and then caudocranially across the FM, and the cumulative caudocranial flow fluctuates below the zero level before the injections. During the lumbar injections, the absorption in the upper cranial SAS leads to a bulk craniocaudal flow of CSF. After the injections, the cumulative caudocranial flow still increases because the ICP has not recovered to the normal level, but the rate of increase decreases. IT injections and the compensatory absorptions lead to net caudocranial CSF flow across the FM. Since experimental measurements of CSF flow rate across FM are scarce, our model can provide a more reliable flow rate boundary condition for partial models like \cite{Putluru2025-dura-thickness} for high-resolution injection simulations. 

\section{Conclusion}\label{sec:conclusion}
%We developed a multiphysics model of the CNS. The model is based on a closed CNS geometry without any artificial boundary condition that is necessary in partial models. In addition to CSF, we model the porous tissue, including the brain parenchyma and spinal cord, and elastic spinal dura mater. 

%We choose the net blood inflow as the driver of the CSF pulsation, avoiding excessive experimental measurements of CSF flow profile. 

We developed a multiphysics model of the CNS that integrates CSF, porous tissue in the brain and spinal cord, and the elastic spinal dura mater. The model is solved within a closed CNS geometry, eliminating the need for uncertain boundary conditions that are required in models that include only parts of the CNS anatomy (e.g., spinal canal or brain). In our model, CSF pulsation is driven by the net blood inflow in the brain, avoiding excessive experimental measurements of CSF flow. The simulation results for physiological conditions indicate that our model captures the craniocaudal decay and phase shift of the CSF flow rate in spinal SAS. Our model also qualitatively captures the caudocranial motion of the spinal cord. In addition, we conduct simulations of intrathecal injections, introducing a source term for injection and a one-way valve sink term for CSF reabsorption. The simulation results show that (1) our model reproduces the elevation and recovery of the ICP during and after the injections; (2) The CSF flow rate across the foramen magnum obtained in injection simulations can serve as a more reliable boundary condition for models of spine subsystems.

Our model has some limitations, which highlight opportunities for further refinements. First, the anatomical microstructures are implicitly modeled by linear momentum contributions, which may underestimate the local CSF flow alternations around the arachnoid trabeculae and overconstrain the CNS tissues. Second, IT injections are modeled as volumetric source terms, which fail to resolve the needle geometry and may not fully capture the local jet flow near the injection location. Third, posture and respiratory effects are not yet included, although they can influence CSF flow pulsations and pressure distribution.

Despite these simplifications, our multiphysics model successfully reproduces key CSF flow features and provides a versatile platform for both physiological and injection investigations, which are essential for advancing intrathecal drug delivery. Our model enables analysis of pathological conditions and therapeutic variations, including differences in anatomy, mechanical properties, and injection strategies such as shunting or decompression. The proposed framework can be extended to introduce additional physiological drivers, such as respiration and cardiac-induced muscular motion. Another interesting topic for future research is building a reliable surrogate model (e.g., via machine learning) to achieve clinical predictions in real time. To conclude, the model enhances our understanding of CSF flow dynamics and provides a platform for clinical predictions and for probing delivery strategies within the CNS. 

\backmatter

%\bmhead{Supplementary information}

%\bmhead{Acknowledgements}

\section*{Declarations}

\begin{itemize}[label={}, leftmargin=0pt]
\item {\textbf{Acknowledgements}}\\
This work is supported by Eli Lilly and company, United States.
\vspace{0.5em}
\item \textbf{Funding} \\
This work was partially supported by Eli Lilly and Company (United States).
\vspace{0.5em}
\item \textbf{Conflict of interest/Competing interests} \\
This research was partially supported by Eli Lilly and Company (United States).
\vspace{0.5em}
\item \textbf{Ethics approval and consent to participate} \\
Not applicable.
\vspace{0.5em}
\item \textbf{Consent for publication} \\
All authors have approved the submission of this manuscript. The content of this
manuscript has not been published or submitted for publication elsewhere.
\vspace{0.5em}
\item \textbf{Data availability}  \\
The data supporting the findings of this study are available within the paper.
\vspace{0.5em}

\item \textbf{Consent to Participate} \\
Not applicable.
\end{itemize}

\begin{appendices}

\section{Net volumetric blood inflow rate waveform}\label{secA-net-blood-inflow}
We adopt the measurement of net blood inflow rate in \cite{Baledent2014-blood-inflow} and fit it with a zero-mean 10-mode Fourier time series, which is expressed as
\begin{align}\label{eq:net-blood-inflow}
    \dot{Q}^{\text{blood}}= 
    \sum_{i=1}^{10} \left[ a_i \cos (i \omega t)
    + b_i \sin (i \omega t) \right].
\end{align}
Here, \( \dot{Q}^{\text{blood}} \) represents the net blood inflow rate, \( a_i\) and \( b_i \) are coefficients of the Fourier series (see Table~\ref{tab:Fourier-coeff}), \( \omega \) is the angular frequency of pulsation, and \( t \in [0,T]\) is the time within a period \(T\) of the cardiac cycle. We assume \( T = 1 \textrm{ s} \), indicating that \( \omega = 2 \pi \).
\begin{table}[h]
    \caption{Fourier coefficients of the net blood inflow rate waveform defined by \eqref{eq:net-blood-inflow}.}\label{tab:Fourier-coeff}%
    \begin{tabular}{@{}rrr@{}}
        \toprule
        \(i\) & \(a_i\) [mL/s] & \(b_i\) [mL/s] \\
        \midrule
        1 & \hphantom{-}1.008215143966994901e+00 & \hphantom{-}3.787753451108464105e+00\\
        2 & -7.521266158330649887e-01 & \hphantom{-}2.232640605779177349e+00 \\
        3 & -8.586527995796720916e-01 & -6.299092534259220884e-01\\
        4 & -7.065404791857486089e-01 & -4.959180340725628184e-01\\
        5 & \hphantom{-}1.154078611129339436e-01 & -5.773618472861138571e-01\\
        6 & \hphantom{-}2.081760182478301999e-01 & \hphantom{-}3.753691557808264179e-02\\
        7 & \hphantom{-}8.410528156592810944e-03 & \hphantom{-}2.553747820668448298e-02\\
        8 & \hphantom{-}3.487466385732233221e-02 & -1.399880099510104969e-02\\
        9 & \hphantom{-}2.907814295856339726e-02 & \hphantom{-}2.452296492753268353e-02\\
        10 & \hphantom{-}1.056115396522858774e-02 & \hphantom{-}6.512830110700565456e-02\\
        \bottomrule
    \end{tabular}
\end{table}

\section{Finite element implementation}\label{secA-FEM}
\setcounter{equation}{0}
In this section, we describe our finite element implementation to solve our multiphysics model. We give the weak forms of the CSF flow equations and the porous tissue mechanics \cite{Ruiz-Baier2022-eye}, as well as the spinal dura mater elastodynamics \cite{Figueroa2006-cmm}. Then we combine them using the interface and transmission conditions \eqref{eq:interface} and \eqref{eq:transmission}, and solve the coupled system monolithically via the direct linear solver {MUMPS} at each time step. We adopt the backward Euler scheme to perform time integration with a time step size \( \Delta t = 0.0125 \) s.

We discretize the computational domain via the continuous Galerkin method. We define the function space \( \mathcal{V} \) for the vector-type variables, for which we use piecewise quadratic elements, and \( \mathcal{W} \) for the scalar-type variables, for which we use piecewise linear elements.

\subsection{Weak form of the unsteady Stokes-Brinkman equations for the CSF flow}\label{secA-FEM-CSF}
The weak form of the unsteady Stokes-Brinkman equations for the CSF flow is: Find \( \bm{u}^f \in \mathcal{V}\) and \( p^f \in \mathcal{W} \), such that for all \( \bm{v}^f \in \mathcal{V} \) and \( q^f \in \mathcal{W} \), 
\begin{align}\label{eq:weak-CSF}
    G^f ={} &
    \int_{\Omega^f} \bm{v}^f \cdot \rho^{f} \frac{\partial \bm{u}^f}{\partial t} \; \mathrm{d}V
    +
    \int_{\Omega^f} \nabla \bm{v}^f : \bm{\sigma}^f \; \mathrm{d}V
    +
    \int_{\Omega^f} \bm{v}^f \cdot \bm{S}^f \; \mathrm{d}V
    -
    \int_{\Omega^f} q^f \nabla \cdot \bm{u}^f \; \mathrm{d}V
    \nonumber \\
    & 
    +
    \int_{\Omega^f} q^f s \; \mathrm{d}V
    -
    \int_{\Sigma} \bm{v}^f \cdot \bm{\sigma}^f \cdot \bm{n}^f \; \mathrm{d}A
    -
    \int_{\Gamma^d} \bm{v}^f \cdot \bm{\sigma}^f \cdot \bm{n}^f \; \mathrm{d}A
    = 0.
\end{align}

\subsection{Weak form of the Biot equations for the brain parenchyma and spinal cord}\label{secA-FEM-porous}
For the robustness of the simulation with nearly-incompressible materials or low storage coefficient, we follow \cite{Ruiz-Baier2022-eye,Causemann2022-fsi-brain} and adopt a three-field formulation \cite{Oyarzua2016-three-field, Lee2017-three-field} of the Biot equations
\begin{subequations}
\begin{alignat}{2}
    - \nabla \cdot \bm{\sigma}^p & = 0, &\quad& \text{in } \Omega^p \times [0,T],
    \label{eq:porous-3-COLM} \\
    \frac{1}{\lambda^p} \phi^{p} - \frac{\alpha}{\lambda^p} p^{p} + \nabla \cdot \bm{d}^p & = 0, &\quad& \text{in } \Omega^p \times [0,T],
    \label{eq:porous-3-total-pressure} \\
    \left( c + \frac{\alpha^2}{\lambda^p} \right) \frac{\partial p^p}{\partial t} 
    - \frac{\alpha}{\lambda^p} \frac{\partial \phi^p}{\partial t} 
    - \nabla \cdot \left( \frac{\kappa}{\mu^f} \nabla p^p  \right)
    & = g, &\quad& \text{in } \Omega^p \times [0,T], \label{eq:porous-3-COM}
\end{alignat}
\end{subequations}
with
\begin{align*}
    \bm{\sigma}^p = -\phi^p \bm{I} 
    + \mu^p \left[ \nabla \bm{d}^p + \left( \nabla \bm{d}^p \right)^T \right].
\end{align*}
Here, \( \phi^p = \alpha p^{p} - \lambda^p \nabla \cdot \bm{d}^p\) is called total pressure. %, which is introduced for robustness.

The weak form of the the Biot equations is: Find \( \bm{d}^p \in \mathcal{V} \), \( p^p \in \mathcal{W} \), and \( \phi^p \in \mathcal{W} \), such that for all \( \bm{w}^p \in \mathcal{V} \), \(q^p \in \mathcal{W}\), and \(\psi^p \in \mathcal{W}\),
\begin{align}\label{eq:weak-porous}
    G^p ={}& \int_{\Omega^p} \nabla \bm{w}^p : \bm{\sigma}^p \; \mathrm{d}V
    - \int_{\Omega^p} \frac{1}{\lambda^p} \psi^p \phi^p \; \mathrm{d}V
    + \int_{\Omega^p} \frac{\alpha}{\lambda^p} \psi^p p^p \; \mathrm{d}V
    - \int_{\Omega^p} \psi^p \nabla \cdot \bm{d}^p \; \mathrm{d}V
    \nonumber \\
    &
    - \int_{\Omega^p} q^p \left( c + \frac{\alpha^2}{\lambda^p} \right) \frac{\partial p^p}{\partial t} \; \mathrm{d}V
    + \int_{\Omega^p} q^p \frac{\alpha}{\lambda^p} \frac{\partial \phi^p}{\partial t} \; \mathrm{d}V
    - \int_{\Omega^p} \nabla q^{p} \cdot \frac{\kappa}{\mu^f} \nabla p^{p} \; \mathrm{d}V
    \nonumber \\
    & 
    + \int_{\Omega^p} q^{p} g \; \mathrm{d}V
    - \int_{\Sigma} \bm{w}^p \cdot \bm{\sigma}^p \cdot \bm{n}^p \; \mathrm{d}A
    + \int_{\Sigma} q^{p} \frac{\kappa}{\mu^f} \nabla p^{p} \cdot \bm{n}^p \; \mathrm{d}A
    = 0.
\end{align}

\subsection{Weak form of the 2D linear elastic equations for the spinal dura mater}\label{secA-FEM-dura}
Since the degrees of freedom of the dura mater and CSF are assumed to be strongly coupled \cite{Figueroa2006-cmm} at the spinal dura mater, we specify that CSF and spinal dura mater share the same velocity field \( \bm{u}^f \) and weight function \( \bm{v}^f \). Then the weak form of the 2D linear elastic equations for the spinal dura mater is: Find \( \bm{u}^f \in \mathcal{V} \), such that for all \( \bm{v}^f \in \mathcal{V}\),
\begin{align}\label{eq:weak-dura}
    G^d = \int_{\Gamma^d} \xi \bm{v}^f \cdot \rho^d \frac{\partial \bm{u}^f}{\partial t} \; \mathrm{d}A + \int_{\Gamma^d} \xi \nabla_{\tau} \bm{v}^f : \bm{\sigma}^d \; \mathrm{d}A - \int_{\Gamma^d} \bm{v}^f \cdot \left( \bm{t}^f + \bm{t}^{\text{pre}} + \bm{t}^e  \right) \; \mathrm{d}A = 0.
\end{align}
Here, we use \( \bm{d}^d (\cdot,t) = \int_{0}^{t}\bm{u}^f(\cdot,\tau)\;\mathrm{d}\tau \) to calculate the displacement of the dura mater, such that there is only the velocity variable in the weak form. Using the backward Euler time scheme, we have \( \bm{d}^d_{n+1} = \bm{d}^d_{n} + \bm{u}^f_{n+1}\Delta t \), where \(n\) and \(n+1\) denote the current and next time step, respectively. 

\subsection{Coupled weak form}\label{secA-FEM-coupled}
Using \eqref{eq:interface} and \eqref{eq:transmission}, we combine \eqref{eq:weak-CSF}, \eqref{eq:weak-porous}, and \eqref{eq:weak-dura} to construct the weak form of the coupled system, which is: Find \( \bm{u}^f \in \mathcal{V}\), \(p^{f} \in \mathcal{W}\), \(\bm{d}^p \in \mathcal{V}\), \( p^p \in \mathcal{W}\), and \( \phi^p \in \mathcal{W} \), such that for all \( \bm{v}^f \in \mathcal{V}\), \(q^{f} \in \mathcal{W}\), \(\bm{w}^p \in \mathcal{V}\), \( q^p \in \mathcal{W}\), and \( \psi^p \in \mathcal{W} \), 
\begin{align}\label{eq:weak-coupled}
    G ={} & G^f + G^p + G^d \nonumber \\
    ={} & 
    \int_{\Omega^f} \bm{v}^f \cdot \rho^{f} \frac{\partial \bm{u}^f}{\partial t} \; \mathrm{d}V
    +
    \int_{\Omega^f} \nabla \bm{v}^f : \bm{\sigma}^f \; \mathrm{d}V
    +
    \int_{\Omega^f} \bm{v}^f \cdot \bm{S}^f \; \mathrm{d}V
    -
    \int_{\Omega^f} q^f \nabla \cdot \bm{u}^f \; \mathrm{d}V
    \nonumber \\
    &
    +\int_{\Omega^f} q^{f}s \;\mathrm{d}V
    + \int_{\Gamma^d} \xi \bm{v}^f \cdot \rho^d \frac{\partial \bm{u}^f}{\partial t} \;\mathrm{d}A + \int_{\Gamma^d} \xi \nabla_{\tau} \bm{v}^f : \bm{\sigma}^d \;\mathrm{d}A - \int_{\Gamma^d} \bm{v}^f \cdot \left( \bm{t}^{\text{pre}} + \bm{t}^e  \right) \; \mathrm{d}A
    \nonumber \\
    &
    +
    \int_{\Sigma} \bm{v}^f \cdot p^p \bm{n}^f \;\mathrm{d}A
    +
    \int_{\Sigma} (\bm{v}^f \times \bm{n}^f) \cdot
    \left[ \left( \bm{u}^f - \frac{\partial \bm{d}^p}{\partial t} \right) \times \bm{n}^f \right] \; \mathrm{d}A
    \nonumber \\
    & + \int_{\Omega^p} \nabla \bm{w}^p : \bm{\sigma}^p \; \mathrm{d}V
    - \int_{\Omega^p} \frac{1}{\lambda^p} \psi^p \phi^p \; \mathrm{d}V
    + \int_{\Omega^p} \frac{\alpha}{\lambda^p} \psi^p p^p \; \mathrm{d}V
    - \int_{\Omega^p} \psi^p \nabla \cdot \bm{d}^p \; \mathrm{d}V
    \nonumber \\
    &
    - \int_{\Omega^p} q^p \left( c + \frac{\alpha^2}{\lambda^p} \right) \frac{\partial p^p}{\partial t} \; \mathrm{d}V
    + \int_{\Omega^p} q^p \frac{\alpha}{\lambda^p} \frac{\partial \phi^p}{\partial t} \; \mathrm{d}V
    - \int_{\Omega^p} \nabla q^{p} \cdot \frac{\kappa}{\mu^f} \nabla p^{p} \; \mathrm{d}V
    \nonumber \\
    & 
    + \int_{\Omega^p} q^{p} g \; \mathrm{d}V
    - \int_{\Sigma} \bm{w}^p \cdot p^{p} \bm{n}^f \; \mathrm{d}A
    - \int_{\Sigma} ( \bm{w}^p \times \bm{n}^f) \cdot \left[ \left( \bm{u}^f - \frac{\partial \bm{d}^p}{\partial t} \right) \times \bm{n}^f \right] \; \mathrm{d}A
    \nonumber \\
    &
    - \int_{\Sigma} \bm{w}^p \cdot \bm{f}^W \; \mathrm{d}A
    + \int_{\Sigma} q^{p} \left( \bm{u}^f - \frac{\partial \bm{d}^p}{\partial t} \right) \cdot \bm{n}^f \; \mathrm{d}A 
    = 0.
\end{align}
In the above equations, we have used the relations \( \bm{n}^p = -\bm{n}^f \) on \( \Sigma \) and  \( 
    \bm{a} \cdot ( \bm{I} -\bm{n}\otimes\bm{n} ) \cdot \bm{b} = (\bm{a}\times\bm{n}) \cdot (\bm{b}\times\bm{n}),
\) where \(\bm{a}\), \(\bm{b}\), and \(\bm{n}\) are vectors.

\subsection{Construction of the unit direction vector of cranial arachonoid trabeculae}\label{secA-FEM-AT}
In Eq.~\eqref{eq:flow-resistance}, we give the expression of the flow resistance term. We use a piecewise constant representation to construct the unit directional vector \( \bm{n}^c \) for the trabeculae. The procedure is as follows:
\begin{enumerate}[label=(\arabic*)]
    \item For elements containing a facet on \( \Gamma^r \), we assign the corresponding facet unit normal vector to the element‐wise unit directional vector \( \bm{n}^c \).
    \item For elements not containing a facet on \(\Gamma^r\), we identify the facet on \(\Gamma^r\) whose centroid is closest to that of the given element and assign its unit normal vector to \( \bm{n}^c \).
\end{enumerate}

\subsection{Mesh generation}\label{secA-FEM-mesh}
We use the open-source meshing software Gmsh \cite{Geuzaine2009-gmsh} to mesh our computational domain with tedrahedral elements.

\end{appendices}


% \bibliography{sn-bibliography}% common bib file
\bibliography{references}

@ARTICLE{Kurtcuoglu2005-idealized-ventricles,
  title    = {Computational modeling of the mechanical behavior of the cerebrospinal fluid system},
  author   = {Kurtcuoglu, Vartan and Poulikakos, Dimos and Ventikos, Yiannis},
  journal  = {Journal of Biomechanical Engineering},
  volume   =  {127},
  number   =  2,
  pages    = "264--269",
  month    =  apr,
  year     =  2005,
  language = "en",
  doi = {10.1115/1.1865191},
}

@ARTICLE{Linge2014-idealized,
  title     = "Effect of craniovertebral decompression on {CSF} dynamics in
               Chiari malformation type {I} studied with computational fluid
               dynamics: Laboratory investigation",
  author    = "Linge, Svein O and Mardal, Kent-A and Helgeland, Anders and
               Heiss, John D and Haughton, Victor",
  journal   = "Journal of Neurosurgery: Spine",
  publisher = "American Association of Neurological Surgeons",
  volume    =  21,
  number    =  4,
  pages     = "559--564",
  month     =  oct,
  year      =  2014,
  language  = "en",
doi = {10.3171/2014.6.SPINE13950}
}

@ARTICLE{Stockman2006-fine-structures,
  title     = "Effect of anatomical fine structure on the flow of cerebrospinal
               fluid in the spinal subarachnoid space",
  author    = "Stockman, Harlan W",
  journal   = "Journal of Biomechanical Engineering",
  publisher = "ASME International",
  volume    =  128,
  number    =  1,
  pages     = "106--114",
  month     =  feb,
  year      =  2006,
  language  = "en",
  doi = {10.1115/1.2132372}    
}

@ARTICLE{Tangen2015-microanatomy,
  title    = "{CNS} wide simulation of flow resistance and drug transport due to
              spinal microanatomy",
  author   = "Tangen, Kevin M and Hsu, Ying and Zhu, David C and Linninger,
              Andreas A",
  journal  = "Journal of Biomechanics",
  volume   =  48,
  number   =  10,
  pages    = "2144--2154",
  month    =  jul,
  year     =  2015,
  language = "en",
doi = {10.1016/j.jbiomech.2015.02.018}
}

@ARTICLE{Heidari-Pahlavian2014-fine-structures,
  title     = "The impact of spinal cord nerve roots and denticulate ligaments
               on cerebrospinal fluid dynamics in the cervical spine",
  author    = "Heidari Pahlavian, Soroush and Yiallourou, Theresia and Tubbs, R
               Shane and Bunck, Alexander C and Loth, Francis and Goodin, Mark
               and Raisee, Mehrdad and Martin, Bryn A",
  journal   = "PLoS One",
  publisher = "Public Library of Science (PLoS)",
  volume    =  9,
  number    =  4,
  pages     = "e91888",
  month     =  apr,
  year      =  2014,
  language  = "en",
doi = {10.1371/journal.pone.0091888}
}

@ARTICLE{Gholampour2018-fsi,
  title     = "{FSI} simulation of {CSF} hydrodynamic changes in a large population of non-communicating hydrocephalus patients during treatment process with regard to their clinical symptoms",
  author    = "Gholampour, Seifollah",
  journal   = "PLoS One",
  publisher = "Public Library of Science",
  volume    =  13,
  number    =  4,
  pages     = "e0196216",
  month     =  apr,
  year      =  2018,
  language  = "en",
doi = {10.1371/journal.pone.0196216}
}

@ARTICLE{Cheng2014-fsi,
  title     = "Effects of fluid structure interaction in a three dimensional
               model of the spinal subarachnoid space",
  author    = "Cheng, Shaokoon and Fletcher, David and Hemley, Sarah and
               Stoodley, Marcus and Bilston, Lynne",
  journal   = "Journal of Biomechanics",
  publisher = "Elsevier BV",
  volume    =  47,
  number    =  11,
  pages     = "2826--2830",
  month     =  aug,
  year      =  2014,
  language  = "en",
doi = {10.1016/j.jbiomech.2014.04.027}
}

@ARTICLE{Linninger2009-fsi-poroelasticity,
  title     = "Normal and hydrocephalic brain dynamics: the role of reduced
               cerebrospinal fluid reabsorption in ventricular enlargement",
  author    = "Linninger, Andreas A and Sweetman, Brian and Penn, Richard",
  journal   = "Annals of Biomedical Engineering ",
  publisher = "Springer Science and Business Media LLC",
  volume    =  37,
  number    =  7,
  pages     = "1434--1447",
  month     =  jul,
  year      =  2009,
  language  = "en",
 doi = {10.1007/s10439-009-9691-4}
}

@ARTICLE{Sweetman2011-hyperelastivity-dura,
  title    = "Cerebrospinal fluid flow dynamics in the central nervous system",
  author   = "Sweetman, Brian and Linninger, Andreas A",
  journal  = "Annals of Biomedical Engineering",
  volume   =  39,
  number   =  1,
  pages    = "484--496",
  month    =  jan,
  year     =  2011,
  language = "en",
 doi = {10.1007/s10439-010-0141-0}
}

@ARTICLE{Khani2017-moving-BC,
  title     = "Nonuniform moving boundary method for computational fluid
               dynamics simulation of intrathecal cerebrospinal flow
               distribution in a Cynomolgus monkey",
  author    = "Khani, Mohammadreza and Xing, Tao and Gibbs, Christina and
               Oshinski, John N and Stewart, Gregory R and Zeller, Jillynne R
               and Martin, Bryn A",
  journal   = "Journal of Biomechanical Engineering",
  publisher = "American Society of Mechanical Engineers",
  volume    =  139,
  number    =  8,
  pages     =  0810051,
  month     =  aug,
  year      =  2017,
  language  = "en",
  doi = {10.1115/1.4036608}
}

@ARTICLE{Howden2008-ventricular-system,
  title     = "Three-dimensional cerebrospinal fluid flow within the human
               ventricular system",
  author    = "Howden, L and Giddings, D and Power, H and Aroussi, A and
               Vloeberghs, M and Garnett, M and Walker, D",
  journal   = "Computer Methods in Biomechanics and Biomedical Engineering ",
  publisher = "Informa UK Limited",
  volume    =  11,
  number    =  2,
  pages     = "123--133",
  month     =  apr,
  year      =  2008,
  language  = "en",
doi = {10.1080/10255840701492118}
}

@ARTICLE{Gupta2010-superior-cSAS,
  title     = "Cerebrospinal fluid dynamics in the human cranial subarachnoid
               space: an overlooked mediator of cerebral disease. {I}.
               Computational model",
  author    = "Gupta, Sumeet and Soellinger, Michaela and Grzybowski, Deborah M
               and Boesiger, Peter and Biddiscombe, John and Poulikakos, Dimos
               and Kurtcuoglu, Vartan",
  journal   = "Journal of the Royal Society Interface",
  publisher = "The Royal Society",
  volume    =  7,
  number    =  49,
  pages     = "1195--1204",
  month     =  aug,
  year      =  2010,
  language  = "en",
  doi = {10.1098/rsif.2010.0033}
}

@ARTICLE{Helgeland2014-cervical-SAS,
  title     = "Numerical simulations of the pulsating flow of cerebrospinal
               fluid flow in the cervical spinal canal of a Chiari patient",
  author    = "Helgeland, Anders and Mardal, Kent-Andre and Haughton, Victor and
               Reif, Bjørn Anders Pettersson",
  journal   = "Journal of Biomechanics",
  publisher = "Elsevier BV",
  volume    =  47,
  number    =  5,
  pages     = "1082--1090",
  month     =  mar,
  year      =  2014,
  language  = "en",
doi={10.1016/j.jbiomech.2013.12.023}
}

@ARTICLE{Yiallourou2012-cervical-SAS,
  title     = "Comparison of {4D} phase-contrast {MRI} flow measurements to
               computational fluid dynamics simulations of cerebrospinal fluid
               motion in the cervical spine",
  author    = "Yiallourou, Theresia I and Kröger, Jan Robert and Stergiopulos,
               Nikolaos and Maintz, David and Martin, Bryn A and Bunck,
               Alexander C",
  journal   = "PLoS One",
  publisher = "Public Library of Science (PLoS)",
  volume    =  7,
  number    =  12,
  pages     = "e52284",
  month     =  dec,
  year      =  2012,
  language  = "en",
doi = {10.1371/journal.pone.0052284}
}

@ARTICLE{Rutkowska2012-cervical-SAS,
  title     = "Patient-specific {3D} simulation of cyclic {CSF} flow at the
               craniocervical region",
  author    = "Rutkowska, G and Haughton, V and Linge, S and Mardal, K-A",
  journal   = "American Journal of Neuroradiology",
  publisher = "American Society of Neuroradiology (ASNR)",
  volume    =  33,
  number    =  9,
  pages     = "1756--1762",
  month     =  oct,
  year      =  2012,
  language  = "en",
  doi = {10.3174/ajnr.A3047}
}

@ARTICLE{Roldan2009-cervical-SAS,
  title     = "Characterization of {CSF} hydrodynamics in the presence and
               absence of tonsillar ectopia by means of computational flow
               analysis",
  author    = "Roldan, A and Wieben, O and Haughton, V and Osswald, T and
               Chesler, N",
  journal   = "American Journal of Neuroradiology",
  publisher = "American Society of Neuroradiology (ASNR)",
  volume    =  30,
  number    =  5,
  pages     = "941--946",
  month     =  may,
  year      =  2009,
  language  = "en",
doi={10.3174/ajnr.A1489}
}

@ARTICLE{Kuttler2010-moving-bc,
  title    = "Understanding pharmacokinetics using realistic computational models of fluid dynamics: biosimulation of drug distribution within the {CSF} space for intrathecal drugs",
  author   = "Kuttler, Andreas and Dimke, Thomas and Kern, Steven and
              Helmlinger, Gabriel and Stanski, Donald and Finelli, Luca A",
  journal  = " Journal of Pharmacokinetics and Pharmacodynamics",
  volume   =  37,
  number   =  6,
  pages    = "629--644",
  month    =  dec,
  year     =  2010,
  language = "en",
doi={10.1007/s10928-010-9184-y}
}

@ARTICLE{Howden2011-full-CNS,
  title     = "Three-dimensional cerebrospinal fluid flow within the human
               central nervous system",
  author    = "Howden, Leo and Giddings, Donald and Power, Henry and Vloeberghs,
               Michael",
  journal   = "Discrete and Continuous Dynamical Systems Series B",
  publisher = "American Institute of Mathematical Sciences (AIMS)",
  volume    =  15,
  number    =  4,
  pages     = "957--969",
  month     =  mar,
  year      =  2011,
  language  = "en",
doi={10.3934/dcdsb.2011.15.957}
}

@ARTICLE{Heidari-Pahlavian2015-drawbacks,
  title     = "Characterization of the discrepancies between four-dimensional
               phase-contrast magnetic resonance imaging and in-silico
               simulations of cerebrospinal fluid dynamics",
  author    = "Heidari Pahlavian, Soroush and Bunck, Alexander C and Loth,
               Francis and Shane Tubbs, R and Yiallourou, Theresia and Kroeger,
               Jan Robert and Heindel, Walter and Martin, Bryn A",
  journal   = "Journal of Biomechanical Engineering",
  publisher = "ASME International",
  volume    =  137,
  number    =  5,
  pages     =  051002,
  month     =  may,
  year      =  2015,
  language  = "en",
  doi = {10.1115/1.4029699}
}

@ARTICLE{Linninger2016-review,
  title     = "Cerebrospinal Fluid Mechanics and Its Coupling to Cerebrovascular
               Dynamics",
  author    = "Linninger, Andreas A and Tangen, Kevin and Hsu, Chih-Yang and
               Frim, David",
  journal   = "Annual Review of Fluid Mechanics",
  publisher = "Annual Reviews",
  volume    =  48,
  number    =  1,
  pages     = "219--257",
  month     =  jan,
  year      =  2016,
  doi = {10.1146/annurev-fluid-122414-034321}
}

@INCOLLECTION{Kurtcuoglu2019-review,
  title     = "Modelling of cerebrospinal fluid flow by computational fluid
               dynamics",
  author    = "Kurtcuoglu, Vartan and Jain, Kartik and Martin, Bryn A",
  booktitle = "Biomechanics of the Brain",
  publisher = "Springer International Publishing",
  address   = "Cham",
  pages     = "215--241",
  series    = "Biological and Medical Physics, Biomedical Engineering",
  year      =  2019,
  language  = "en",
 doi = {10.1007/978-3-030-04996-6_9}
}

@book{Cushing1926-MK-doctrine,
  author    = {Cushing, Harvey},
  title     = {Studies in intracranial physiology \& surgery; the third circulation, the hypophysics, the gliomas},
  publisher = {H. Milford, Oxford University Press},
  address   = {London, United Kingdom},
  year      = {1926},
  language  = {en}
}

@ARTICLE{Johanson2008-CSF-production,
  title     = "Multiplicity of cerebrospinal fluid functions: New challenges in health and disease",
  author    = "Johanson, Conrad E and Duncan, III, John A and Klinge, Petra M and Brinker, Thomas and Stopa, Edward G and Silverberg, Gerald D",
  journal   = "Cerebrospinal Fluid Research",
  publisher = "Springer Science and Business Media LLC",
  volume    =  5,
  number    =  1,
  pages     =  10,
  month     =  may,
  year      =  2008,
  language  = "en",
doi={10.1186/1743-8454-5-10}
}

@ARTICLE{Bhadelia1997-CSF-pulsation,
  title     = "Cerebrospinal fluid pulsation amplitude and its quantitative
               relationship to cerebral blood flow pulsations: a phase-contrast
               {MR} flow imaging study",
  author    = "Bhadelia, R A and Bogdan, A R and Kaplan, R F and Wolpert, S M",
  journal   = "Neuroradiology",
  publisher = "Springer Science and Business Media LLC",
  volume    =  39,
  number    =  4,
  pages     = "258--264",
  month     =  apr,
  year      =  1997,
  language  = "en",
doi={10.1007/s002340050404}
}

@ARTICLE{Greitz2004-CSF,
  title     = "Radiological assessment of hydrocephalus: new theories and
               implications for therapy",
  author    = "Greitz, Dan",
  journal   = "Neurosurgical Review",
  publisher = "Springer Science and Business Media LLC",
  volume    =  27,
  number    =  3,
  pages     = "145--65; discussion 166--7",
  month     =  jul,
  year      =  2004,
  language  = "en",
doi={10.1007/s10143-004-0326-9}
}

@ARTICLE{Daouk2017-cardiac-driven,
  title     = "Heart rate and respiration influence on macroscopic blood and
               {CSF} flows",
  author    = "Daouk, Joël and Bouzerar, Roger and Baledent, Olivier",
  journal   = "Acta Radiologica",
  publisher = "SAGE Publications",
  volume    =  58,
  number    =  8,
  pages     = "977--982",
  month     =  aug,
  year      =  2017,
  language  = "en",
doi={10.1177/0284185116676655}
}

@ARTICLE{Haughton2014-cardiac-driven,
  title     = "Spinal fluid biomechanics and imaging: an update for
               neuroradiologists",
  author    = "Haughton, V and Mardal, K-A",
  journal   = "American Journal of Neuroradiology",
  publisher = "American Society of Neuroradiology (ASNR)",
  volume    =  35,
  number    =  10,
  pages     = "1864--1869",
  month     =  oct,
  year      =  2014,
  language  = "en",
doi={10.3174/ajnr.A4023}
}

@ARTICLE{Eklund2007-RC,
  title    = "Assessment of cerebrospinal fluid outflow resistance",
  author   = "Eklund, Anders and Smielewski, Peter and Chambers, Iain and
              Alperin, Noam and Malm, Jan and Czosnyka, Marek and Marmarou,
              Anthony",
  journal  = "Medical \& Biological Engineering \& Computing",
  volume   =  45,
  number   =  8,
  pages    = "719--735",
  month    =  aug,
  year     =  2007,
  language = "en",
doi={10.1007/s11517-007-0199-5}
}

@ARTICLE{Raboel2012-review-ICP-measurement,
  title     = "Intracranial pressure monitoring: Invasive versus non-invasive
               methods-A review",
  author    = "Raboel, P H and Bartek, Jr, J and Andresen, M and Bellander, B M and Romner, B",
  journal   = "Critical Care Research and Practice",
  publisher = "Hindawi Limited",
  volume    =  2012,
  pages     =  950393,
  month     =  jun,
  year      =  2012,
  language  = "en",
  doi = {10.1155/2012/950393}
}

@ARTICLE{Rajajee2024-noninvasive,
  title     = "Noninvasive intracranial pressure monitoring: Are we there yet?",
  author    = "Rajajee, Venkatakrishna",
  journal   = "Neurocritical Care",
  publisher = "Springer Science and Business Media LLC",
  volume    =  41,
  number    =  2,
  pages     = "332--338",
  month     =  oct,
  year      =  2024,
  language  = "en",
  doi = {10.1007/s12028-024-01951-1}
}

@ARTICLE{Proulx2021-CSF-absorption,
  title     = "Cerebrospinal fluid outflow: a review of the historical and
               contemporary evidence for arachnoid villi, perineural routes, and
               dural lymphatics",
  author    = "Proulx, Steven T",
  journal   = "Cellular and Molecular Life Sciences",
  publisher = "Springer Science and Business Media LLC",
  volume    =  78,
  number    =  6,
  pages     = "2429--2457",
  month     =  mar,
  year      =  2021,
  language  = "en",
doi={10.1007/s00018-020-03706-5}
}

@article{Causemann2022-fsi-brain,
  author    = {Causemann, Marius and Vinje, Vegard and Rognes, Marie E.},
  title     = {Human intracranial pulsatility during the cardiac cycle: a computational modelling framework},
  journal   = {Fluids and Barriers of the CNS},
  volume    = {19},
  number    = {1},
  pages     = {84},
  year      = {2022},
  doi       = {10.1186/s12987-022-00376-2}
}

@article{Ruiz-Baier2022-eye,
  author    = {Ruiz-Baier, Ricardo and Taffetani, Matteo and Westermeyer, Hans D. and Yotov, Ivan},
  title     = {The {Biot–Stokes} coupling using total pressure: Formulation, analysis and application to interfacial flow in the eye},
  journal   = {Computer Methods in Applied Mechanics and Engineering},
  volume    = {389},
  pages     = {114384},
  year      = {2022},
  issn      = {0045-7825},
  doi       = {10.1016/j.cma.2021.114384}
}

@article{Biot1941-Biot-theory,
  author    = {Biot, Maurice A.},
  title     = {General theory of three-dimensional consolidation},
  journal   = {Journal of Applied Physics},
  volume    = {12},
  number    = {2},
  pages     = {155--164},
  month     = {Feb},
  year      = {1941},
  doi       = {10.1063/1.1712886}
}

@article{Oyarzua2016-three-field,
  author    = {Oyarzúa, Ricardo and Ruiz-Baier, Ricardo},
  title     = {Locking-Free Finite Element Methods for Poroelasticity},
  journal   = {SIAM Journal on Numerical Analysis},
  month     = {Sep},
volume = {54},
number = {5},
pages = {2951-2973},
  year      = {2016},
  doi       = {10.1137/15M1050082}
}

@article{Lee2017-three-field,
  author    = {Lee, Jeonghun J. and Mardal, Kent-Andre and Winther, Ragnar},
  title     = {Parameter-Robust Discretization and Preconditioning of {B}iot's Consolidation Model},
  journal   = {SIAM Journal on Scientific Computing},
  month     = {Jan},
  year      = {2017},
    volume = {39},
    number = {1},
    pages = {A1-A24},
  doi       = {10.1137/15M1029473}
}

@ARTICLE{Saffman1971-BJF,
  title     = "On the boundary condition at the surface of a porous medium",
  author    = "Saffman, P G",
  journal   = "Studies in Applied Mathematics",
  publisher = "Wiley",
  volume    =  50,
  number    =  2,
  pages     = "93--101",
  month     =  jun,
  year      =  1971,
  language  = "en",
doi={10.1002/sapm197150293}
}

@ARTICLE{Beavers1967-BJF,
  title     = "Boundary conditions at a naturally permeable wall",
  author    = "Beavers, Gordon S and Joseph, Daniel D",
  journal   = "Journal of Fluid Mechanics",
  publisher = "Cambridge University Press (CUP)",
  volume    =  30,
  number    =  1,
  pages     = "197--207",
  month     =  oct,
  year      =  1967,
  language  = "en",
doi = {10.1017/S0022112067001375}
}

@article{Gupta2009-flow-resistance,
  author    = {Gupta, Sumeet and Soellinger, Michaela and Boesiger, Peter and Poulikakos, Dimos and Kurtcuoglu, Vartan},
  title     = {Three-dimensional computational modeling of subject-specific cerebrospinal fluid flow in the subarachnoid space},
  journal   = {Journal of Biomechanical Engineering},
  volume    = {131},
  number    = {2},
  pages     = {021010},
  month     = {Feb},
  year      = {2009},
  issn      = {0148-0731},
  doi       = {10.1115/1.3005171}
}

@article{Brinkman1949-porous,
  author    = {Brinkman, H.~C.},
  title     = {On the permeability of media consisting of closely packed porous particles},
  journal   = {Applied Scientific Research},
  volume    = {1},
  number    = {1},
  pages     = {81--86},
  month     = {Dec},
  year      = {1949},
  doi       = {10.1007/BF02120318}
}

@article{Westhuizen1994-flow-resistance,
  author    = {Westhuizen, Josiasvd and Plessis, J. Prieur Du},
  title     = {Quantification of unidirectional fiber bed permeability},
  journal   = {Journal of Composite Materials},
  publisher = {SAGE Publications},
  volume    = {28},
  number    = {7},
  pages     = {619--637},
  month     = {May},
  year      = {1994},
  doi       = {10.1177/002199839402800703},
  language  = {en}
}

@article{Figueroa2006-cmm,
  author    = {Figueroa, C. Alberto and Vignon-Clementel, Irene E. and Jansen, Kenneth E. and Hughes, Thomas J.R. and Taylor, Charles A.},
  title     = {A coupled momentum method for modeling blood flow in three-dimensional deformable arteries},
  journal   = {Computer Methods in Applied Mechanics and Engineering},
  volume    = {195},
  number    = {41},
  pages     = {5685--5706},
  year      = {2006},
  issn      = {0045-7825},
  doi       = {10.1016/j.cma.2005.11.011},
}

@book{Hughes2000-fembook,
  author    = {Hughes, Thomas J.R.},
  title     = {The Finite Element Method: Linear Static and Dynamic Finite Element Analysis},
  publisher = {Dover Publications},
  address   = {Mineola, New York, USA},
  year      = {2000}
}

@article{Moireau2012-ext-tiss-sup,
  author    = {Moireau, Philippe and Xiao, Nan and Astorino, Matteo and Figueroa, C. Alberto and Chapelle, Dominique and Taylor, Charles A. and Gerbeau, J.-F.},
  title     = {External tissue support and fluid-structure simulation in blood flows},
  journal   = {Biomechanics and Modeling in Mechanobiology},
  volume    = {11},
  number    = "1-2",
  pages     = {1--18},
  year      = {2012},
  publisher = {Springer},
  doi       = {10.1016/j.cma.2005.11.011}
}

@incollection{Baledent2014-blood-inflow,
  author    = {Balédent, Olivier},
  title     = {Imaging of the cerebrospinal fluid circulation},
  booktitle = {Adult Hydrocephalus},
  editor    = {Rigamonti, Daniele},
  pages     = {121--138},
  publisher = {Cambridge University Press},
  address   = {Cambridge, United Kingdom},
  year      = {2014},
  doi       = {10.1017/CBO9781139382816.013}
}

@article{Sass2017-spine-geometry,
  author    = {Sass, Lucas R. and Khani, Mohammadreza and Natividad, Gabryel Connely and Tubbs, R. Shane and Balédent, Olivier and Martin, Bryn A.},
  title     = {A 3{D} subject-specific model of the spinal subarachnoid space with anatomically realistic ventral and dorsal spinal cord nerve rootlets},
  journal   = {Fluids and Barriers of the CNS},
  volume    = {14},
  number    = {1},
  pages     = {36},
  month     = {Dec},
  year      = {2017},
  doi       = {10.1186/s12987-017-0085-y}
}

@ARTICLE{Ekstedt1978-absorption-coeff,
  title     = "{CSF} hydrodynamic studies in man. 2 . Normal hydrodynamic
               variables related to {CSF} pressure and flow",
  author    = "Ekstedt, J",
  journal   = "Journal of Neurology, Neurosurgery and Psychiatry",
  publisher = "BMJ",
  volume    =  41,
  number    =  4,
  pages     = "345--353",
  month     =  apr,
  year      =  1978,
  language  = "en",
  doi = {10.1136/jnnp.41.4.345},
}

@ARTICLE{Runza1999-dura-elasticity,
  title     = "Lumbar Dura mater biomechanics: Experimental characterization and
               scanning electron microscopy observations",
  author    = "Runza, Massimo and Pietrabissa, Riccardo and Mantero, Sara and
               Albani, Alessandro and Quaglini, Virginio and Contro, Roberto",
  journal   = "Anesthesia \& Analgesia ",
  publisher = "Ovid Technologies (Wolters Kluwer Health)",
  volume    =  88,
  number    =  6,
  pages     = "1317--1321",
  month     =  jun,
  year      =  1999,
  language  = "en",
  doi = {10.1213/00000539-199906000-00022},
}

@article{Mcintosh2010-tissue-param,
  author    = {Mcintosh, Robert L. and Anderson, Vitas},
  title     = {A comprehensive tissue properties database provided for the thermal assessment of a human at rest},
  journal   = {Biophysical Reviews and Letters},
  volume    = {5},
  number    = {3},
  pages     = {129--151},
  year      = {2010},
  doi       = {10.1142/S1793048010001184}
}

@article{Benko2020-AT-distribution,
  author    = {Benko, Nikolaus and Luke, Emma and Alsanea, Yousef and Coats, Brittany},
  title     = {Spatial distribution of human arachnoid trabeculae},
  journal   = {Journal of Anatomy},
  volume    = {237},
  number    = {2},
  pages     = {275--284},
  month     = {Aug},
  year      = {2020},
  doi       = {10.1111/joa.13186}
}

@article{Albeck1991-absorption-coeff,
  author    = {Albeck, M.~J. and Børgesen, S.~E. and Gjerris, F. and Schmidt, J.~F. and Sørensen, P.~S.},
  title     = {Intracranial pressure and cerebrospinal fluid outflow conductance in healthy subjects},
  journal   = {Journal of Neurosurgery},
  volume    = {74},
  number    = {4},
  pages     = {597--600},
  month     = {Apr},
  year      = {1991},
  doi       = {10.3171/jns.1991.74.4.0597}
}

@ARTICLE{Alkhouli2013-fat-modulus,
  title     = "The mechanical properties of human adipose tissues and their
               relationships to the structure and composition of the
               extracellular matrix",
  author    = "Alkhouli, Nadia and Mansfield, Jessica and Green, Ellen and Bell,
               James and Knight, Beatrice and Liversedge, Neil and Tham, Ji
               Chung and Welbourn, Richard and Shore, Angela C and Kos, Katarina
               and Winlove, C Peter",
  journal   = "American Journal of Physiology-Endocrinology and Metabolism
",
  publisher = "American Physiological Society",
  volume    =  305,
  number    =  12,
  pages     = "E1427--35",
  month     =  dec,
  year      =  2013,
  language  = "en",
doi={10.1152/ajpendo.00111.2013}
}

@ARTICLE{Benko2021-trabeculae-modulus,
  title     = "Mechanical characterization of the human pia-arachnoid complex",
  author    = "Benko, Nikolaus and Luke, Emma and Alsanea, Yousef and Coats,
               Brittany",
  journal   = "Journal of the Mechanical Behavior of Biomedical Materials",
  publisher = "Elsevier BV",
  volume    =  120,
  number    =  104579,
  pages     =  104579,
  month     =  aug,
  year      =  2021,
  language  = "en",
doi={10.1016/j.jmbbm.2021.104579}
}

@ARTICLE{Cavelier2022-dura-properties,
  title     = "Tensile properties of human spinal dura mater and pericranium",
  author    = "Cavelier, Sacha and Quarrington, Ryan D and Jones, Claire F",
  journal   = "Journal of Materials Science: Materials in Medicine ",
  publisher = "Springer Science and Business Media LLC",
  volume    =  34,
  number    =  1,
  pages     =  4,
  month     =  dec,
  year      =  2022,
  language  = "en",
doi={10.1007/s10856-022-06704-0}
}

@ARTICLE{Rangel-Castilla2008-pressure-range,
  title     = "Management of intracranial hypertension",
  author    = "Rangel-Castilla, Leonardo and Gopinath, Shankar and Robertson,
               Claudia S",
  journal   = "Neurologic Clinics",
  publisher = "Elsevier BV",
  volume    =  26,
  number    =  2,
  pages     = "521--541",
  month     =  may,
  year      =  2008,
  language  = "en",
doi={10.1016/j.ncl.2008.02.003}
}

@ARTICLE{Holter2017-permeability,
  title     = "Interstitial solute transport in {3D} reconstructed neuropil
               occurs by diffusion rather than bulk flow",
  author    = "Holter, Karl Erik and Kehlet, Benjamin and Devor, Anna and
               Sejnowski, Terrence J and Dale, Anders M and Omholt, Stig W and
               Ottersen, Ole Petter and Nagelhus, Erlend Arnulf and Mardal,
               Kent-André and Pettersen, Klas H",
  journal   = "Proceedings of the National Academy of Sciences of the United States of America",
  publisher = "National Academy of Sciences",
  volume    =  114,
  number    =  37,
  pages     = "9894--9899",
  month     =  sep,
  year      =  2017,
  language  = "en",
doi={10.1073/pnas.1706942114}
}

@ARTICLE{Smith2007-brain-tissue,
  title     = "Interstitial transport and transvascular fluid exchange during
               infusion into brain and tumor tissue",
  author    = "Smith, Joshua H and Humphrey, Joseph A C",
  journal   = "Microvascular Research",
  publisher = "Elsevier BV",
  volume    =  73,
  number    =  1,
  pages     = "58--73",
  month     =  jan,
  year      =  2007,
  language  = "en",
doi={10.1016/j.mvr.2006.07.001}
}

@ARTICLE{Barber1970-density,
  title     = "The density of tissues in and about the head",
  author    = "Barber, T W and Brockway, J A and Higgins, L S",
  journal   = "Acta Neurologica Scandinavica",
  publisher = "Hindawi Limited",
  volume    =  46,
  number    =  1,
  pages     = "85--92",
  month     =  mar,
  year      =  1970,
  language  = "en",
doi={10.1111/j.1600-0404.1970.tb05606.x}
}

@incollection{Ehrhardt2012-slip-rate-coeff,
    author = {Ehrhardt, Matthias},
    year = {2012},
    month = {07},
    pages = {3-12},
    title = {An Introduction to Fluid-Porous Interface Coupling},
    booktitle = {Progress in Computational Physics (PiCP): Coupled Fluid Flow in Energy, Biology and Environmental Research},
    volume = {2},
    isbn = {9781608052547},
    publisher = {Bentham Science Publisher},
    address = {Sharjah, UAE},
    doi = {10.2174/978160805254711202010003}
}

@ARTICLE{Bloomfield1998-viscosity,
  title     = "Effects of proteins, blood cells and glucose on the viscosity of
               cerebrospinal fluid",
  author    = "Bloomfield, I G and Johnston, I H and Bilston, L E",
  journal   = "Pediatric Neurosurgery",
  publisher = "S. Karger AG",
  volume    =  28,
  number    =  5,
  pages     = "246--251",
  month     =  may,
  year      =  1998,
  language  = "en",
doi={10.1159/000028659}
}

@ARTICLE{Budday2015-Young-modulus,
  title     = "Mechanical properties of gray and white matter brain tissue by
               indentation",
  author    = "Budday, Silvia and Nay, Richard and de Rooij, Rijk and Steinmann,
               Paul and Wyrobek, Thomas and Ovaert, Timothy C and Kuhl, Ellen",
  journal   = "Journal of the Mechanical Behavior of Biomedical Materials",
  publisher = "Elsevier BV",
  volume    =  46,
  pages     = "318--330",
  month     =  jun,
  year      =  2015,
  language  = "en",
doi={10.1016/j.jmbbm.2015.02.024}
}

@ARTICLE{Chou2016-storage-coeff,
  title     = "A fully dynamic multi-compartmental poroelastic system:
               Application to aqueductal stenosis",
  author    = "Chou, Dean and Vardakis, John C and Guo, Liwei and Tully, Brett J
               and Ventikos, Yiannis",
  journal   = "Journal of Biomechanics",
  publisher = "Elsevier",
  volume    =  49,
  number    =  11,
  pages     = "2306--2312",
  month     =  jul,
  year      =  2016,
  language  = "en",
doi={10.1016/j.jbiomech.2015.11.025}
}

@ARTICLE{Guo2019-storage-coeff,
  title     = "On the validation of a multiple-network poroelastic model using
               arterial spin labeling {MRI} data",
  author    = "Guo, Liwei and Li, Zeyan and Lyu, Jinhao and Mei, Yuqian and
               Vardakis, John C and Chen, Duanduan and Han, Cong and Lou, Xin
               and Ventikos, Yiannis",
  journal   = "Frontiers in Computational Neuroscience",
  publisher = "Frontiers Media SA",
  volume    =  13,
  pages     =  60,
  month     =  sep,
  year      =  2019,
  language  = "en",
doi={10.3389/fncom.2019.00060}
}

@ARTICLE{Putluru2025-dura-thickness,
  title     = "Mixed-dimensional fluid–structure interaction simulations reveal
               key mechanisms of cerebrospinal fluid dynamics in the spinal
               canal",
  author    = "Putluru, Deshik Reddy and Tepole, Adrian Buganza and Gomez,
               Hector",
  journal   = "Fluids and Barriers of the CNS",
  publisher = "Springer Science and Business Media LLC",
  volume    =  22,
  number    =  1,
  pages     = "1--20",
  month     =  jul,
  year      =  2025,
  language  = "en",
  doi = {10.1186/s12987-025-00691-4}
}

@article{AlnaesEtal2015-fenics,
  author    = {Aln{\ae}s, Martin S. and Blechta, Jan and Hake, Johan and Johansson, August and Kehlet, Benjamin and Logg, Anders and Richardson, Chris N. and Ring, Johannes and Rognes, Marie E. and Wells, Garth N.},
  title     = {The {FEniCS} Project Version 1.5},
  journal   = {Archive of Numerical Software},
  volume    = {3},
  year      = {2015},
  doi       = {10.11588/ans.2015.100.20553}
}

@misc{multiphenics,
  author    = {Ballarin, Francesco},
  title     = {\href{https://mathlab.sissa.it/multiphenics}{Multiphenics – easy prototyping of multiphysics problems in FEniCS}},
  year      = {2020},
  note      = {Accessed: Nov 8, 2024}
}

@article{Geuzaine2009-gmsh,
  author    = {Geuzaine, Christophe and Remacle, Jean-François},
  title     = {{Gmsh: A 3-D finite element mesh generator with built-in pre- and post-processing facilities}},
  journal   = {International Journal for Numerical Methods in Engineering},
  volume    = {79},
  number    = {11},
  pages     = {1309--1331},
  year      = {2009},
  doi       = {10.1002/nme.2579}
}

@book{Kandel2013-CNS-function,
  title = {Principles of neural science},
  author = {Kandel, Eric R and Schwartz, James H and Jessell, Thomas M and Siegelbaum, Steven and Hudspeth, A J},
  edition = {5th},
  year = {2013},
  publisher = {McGraw‑Hill Education},
  address = {New York, NY, USA},
}

@book{WHO2006-neurological,
  title     = {Neurological Disorders: Public Health Challenges},
  author    = {{World Health Organization}},
  year      = {2006},
  publisher = {World Health Organization},
  address   = {Geneva, Switzerland},
}

@ARTICLE{Khani2020-in-vitro,
  title     = "In vitro and numerical simulation of blood removal from
               cerebrospinal fluid: comparison of lumbar drain to Neurapheresis
               therapy",
  author    = "Khani, Mohammadreza and Sass, Lucas R and Sharp, M Keith and
               McCabe, Aaron R and Zitella Verbick, Laura M and Lad, Shivanand P
               and Martin, Bryn A",
  journal   = "Fluids and Barriers of the CNS",
  publisher = "Springer Science and Business Media LLC",
  volume    =  17,
  number    =  1,
  pages     =  23,
  month     =  mar,
  year      =  2020,
  language  = "en",
doi = {10.1186/s12987-020-00185-5}
}

@ARTICLE{Tangen2017-in-vitro,
  title    = "Computational and In Vitro Experimental Investigation of
              Intrathecal Drug Distribution: Parametric Study of the Effect of
              Injection Volume, Cerebrospinal Fluid Pulsatility, and Drug Uptake",
  author   = "Tangen, Kevin M and Leval, Roxanne and Mehta, Ankit I and
              Linninger, Andreas A",
  journal  = "Anesthesia \& Analgesia",
  volume   =  124,
  number   =  5,
  pages    = "1686--1696",
  month    =  may,
  year     =  2017,
  language = "en",
doi = {10.1213/ANE.0000000000002011}
}

@ARTICLE{Liu2022-in-vivo,
  title     = "Measurements of cerebrospinal fluid production: a review of the limitations and advantages of current methodologies",
  author    = "Liu, Guojun and Ladrón-de-Guevara, Antonio and Izhiman, Yara and
               Nedergaard, Maiken and Du, Ting",
  journal   = "Fluids and Barriers of the CNS",
  publisher = "BioMed Central",
  volume    =  19,
  number    =  1,
  pages     =  101,
  month     =  dec,
  year      =  2022,
  language  = "en",
doi = {10.1186/s12987-022-00382-4}
}

@ARTICLE{Rivera-Rivera2024-MRI-review,
  title     = "Four-dimensional flow {MRI} for quantitative assessment of
               cerebrospinal fluid dynamics: Status and opportunities",
  author    = "Rivera-Rivera, Leonardo A and Vikner, Tomas and Eisenmenger,
               Laura and Johnson, Sterling C and Johnson, Kevin M",
  journal   = "NMR in Biomedicine",
  publisher = "Wiley",
  volume    =  37,
  number    =  7,
  pages     = "e5082",
  month     =  jul,
  year      =  2024,
  language  = "en",
doi = {10.1002/nbm.5082}
}

@incollection{Tangen2018-IT-review,
title = {Cerebrospinal Fluid Dynamics and Intrathecal Delivery},
author = {Tangen, Kevin and Mehta, Ankit I. and Linninger, Andreas A.},
editor = {Elliot S. Krames and P. Hunter Peckham and Ali R. Rezai},
booktitle = {Neuromodulation},
edition   = {second},
publisher = {Academic Press, Elsevier},
address = {London, United Kingdom},
pages = {829--846},
year = {2018},
isbn = {978-0-12-805353-9},
doi = {10.1016/B978-0-12-805353-9.00067-X},
}

@ARTICLE{De-Andres2022-IT-review,
  title    = "Intrathecal Drug Delivery: Advances and Applications in the Management of Chronic Pain Patient",
  author   = "De Andres, Jose and Hayek, Salim and Perruchoud, Christophe and Lawrence, Melinda M and Reina, Miguel Angel and De Andres-Serrano, Carmen and Rubio-Haro, Ruben and Hunt, Mathew and Yaksh, Tony L",
  journal  = "Frontiers in Pain Research",
  volume   =  3,
  pages    =  900566,
  month    =  jun,
  year     =  2022,
  language = "en",
 doi = {10.3389/fpain.2022.900566}
}

@ARTICLE{Mazur2019-IT-review,
  title     = "Brain pharmacology of intrathecal antisense oligonucleotides revealed through multimodal imaging",
  author    = "Mazur, Curt and Powers, Berit and Zasadny, Kenneth and Sullivan, Jenna M and Dimant, Hemi and Kamme, Fredrik and Hesterman, Jacob
               and Matson, John and Oestergaard, Michael and Seaman, Marc and
               Holt, Robert W and Qutaish, Mohammed and Polyak, Ildiko and
               Coelho, Richard and Gottumukkala, Vijay and Gaut, Carolynn M and Berridge, Marc and Albargothy, Nazira J and Kelly, Louise and
               Carare, Roxana O and Hoppin, Jack and Kordasiewicz, Holly and
               Swayze, Eric E and Verma, Ajay",
  journal   = "JCI Insight",
  publisher = "The American Society for Clinical Investigation",
  volume    =  4,
  number    =  20,
  pages     = {{e129240}},
  month     =  oct,
  year      =  2019,
  language  = "en",
  doi = {10.1172/jci.insight.129240},
}

@ARTICLE{Belov2021-experiments-injection,
  title    = "Large-Volume Intrathecal Administrations: Impact on {CSF} Pressure
              and Safety Implications",
  author   = "Belov, Vasily and Appleton, Janine and Levin, Stepan and Giffenig,
              Pilar and Durcanova, Beata and Papisov, Mikhail",
  journal  = "Frontiers in Neuroscience",
  volume   =  15,
  pages    =  604197,
  month    =  apr,
  year     =  2021,
  language = "en",
doi={10.3389/fnins.2021.604197}
}

@ARTICLE{Coenen2019-spinal-SAS-experiment,
  title     = "Subject-specific studies of {CSF} bulk flow patterns in the
               spinal canal: Implications for the dispersion of solute particles
               in intrathecal drug delivery",
  author    = "Coenen, W and Gutiérrez-Montes, C and Sincomb, S and
               Criado-Hidalgo, E and Wei, K and King, K and Haughton, V and
               Martínez-Bazán, C and Sánchez, A L and Lasheras, J C",
  journal   = "American Journal of Neuroradiology",
  publisher = "American Society of Neuroradiology (ASNR)",
  volume    =  40,
  number    =  7,
  pages     = "1242--1249",
  month     =  jul,
  year      =  2019,
  language  = "en",
doi={10.3174/ajnr.A6097}
}

@ARTICLE{Wolf2023-sc-displacement,
  title     = "{CSF} flow and spinal cord motion in patients with spontaneous
               intracranial hypotension: A phase contrast {MRI} study",
  author    = "Wolf, Katharina and Luetzen, Niklas and Mast, Hansjoerg and
               Kremers, Nico and Reisert, Marco and Beltrán, Saúl and Fung,
               Christian and Beck, Jürgen and Urbach, Horst",
  journal   = "Neurology",
  publisher = "Ovid Technologies (Wolters Kluwer Health)",
  volume    =  100,
  number    =  7,
  pages     = "e651--e660",
  month     =  feb,
  year      =  2023,
  language  = "en",
doi={10.1212/WNL.0000000000201527}
}

@article{DeAndres2011-epidural-fat-distribution,
  author = {De Andrés, J. and Reina, M. A. and Machés, F. and De Sola, R. G. and Oliva, A. and Prats-Galino, A.},
  title = {Epidural fat: considerations for minimally invasive spinal injection and surgical therapies},
  journal = {Journal of Neurosurgical Review},
  year = {2011},
  volume = {1},
  pages = {45--53}
}

\end{document}